\documentclass[nolinenumbers]{aastex631}

\usepackage{amsmath}
\usepackage{float}
\usepackage{comment}
\usepackage{array}
\newcommand{\PreserveBackslash}[1]{\let\temp=\\#1\let\\=\temp}
\newcolumntype{C}[1]{>{\PreserveBackslash\centering}p{#1}}
\usepackage[T5,T1]{fontenc}
\graphicspath{{./}{figures/}}

\begin{document}

\title{Searches for Neutrinos from Gamma-Ray Bursts using the IceCube Neutrino Observatory}

\affiliation{III. Physikalisches Institut, RWTH Aachen University, D-52056 Aachen, Germany}
\affiliation{Department of Physics, University of Adelaide, Adelaide, 5005, Australia}
\affiliation{Dept. of Physics and Astronomy, University of Alaska Anchorage, 3211 Providence Dr., Anchorage, AK 99508, USA}
\affiliation{Dept. of Physics, University of Texas at Arlington, 502 Yates St., Science Hall Rm 108, Box 19059, Arlington, TX 76019, USA}
\affiliation{CTSPS, Clark-Atlanta University, Atlanta, GA 30314, USA}
\affiliation{School of Physics and Center for Relativistic Astrophysics, Georgia Institute of Technology, Atlanta, GA 30332, USA}
\affiliation{Dept. of Physics, Southern University, Baton Rouge, LA 70813, USA}
\affiliation{Dept. of Physics, University of California, Berkeley, CA 94720, USA}
\affiliation{Lawrence Berkeley National Laboratory, Berkeley, CA 94720, USA}
\affiliation{Institut f{\"u}r Physik, Humboldt-Universit{\"a}t zu Berlin, D-12489 Berlin, Germany}
\affiliation{Fakult{\"a}t f{\"u}r Physik {\&} Astronomie, Ruhr-Universit{\"a}t Bochum, D-44780 Bochum, Germany}
\affiliation{Universit{\'e} Libre de Bruxelles, Science Faculty CP230, B-1050 Brussels, Belgium}
\affiliation{Vrije Universiteit Brussel (VUB), Dienst ELEM, B-1050 Brussels, Belgium}
\affiliation{Department of Physics and Laboratory for Particle Physics and Cosmology, Harvard University, Cambridge, MA 02138, USA}
\affiliation{Dept. of Physics, Massachusetts Institute of Technology, Cambridge, MA 02139, USA}
\affiliation{Dept. of Physics and The International Center for Hadron Astrophysics, Chiba University, Chiba 263-8522, Japan}
\affiliation{Department of Physics, Loyola University Chicago, Chicago, IL 60660, USA}
\affiliation{Dept. of Physics and Astronomy, University of Canterbury, Private Bag 4800, Christchurch, New Zealand}
\affiliation{Dept. of Physics, University of Maryland, College Park, MD 20742, USA}
\affiliation{Dept. of Astronomy, Ohio State University, Columbus, OH 43210, USA}
\affiliation{Dept. of Physics and Center for Cosmology and Astro-Particle Physics, Ohio State University, Columbus, OH 43210, USA}
\affiliation{Niels Bohr Institute, University of Copenhagen, DK-2100 Copenhagen, Denmark}
\affiliation{Dept. of Physics, TU Dortmund University, D-44221 Dortmund, Germany}
\affiliation{Dept. of Physics and Astronomy, Michigan State University, East Lansing, MI 48824, USA}
\affiliation{Dept. of Physics, University of Alberta, Edmonton, Alberta, Canada T6G 2E1}
\affiliation{Erlangen Centre for Astroparticle Physics, Friedrich-Alexander-Universit{\"a}t Erlangen-N{\"u}rnberg, D-91058 Erlangen, Germany}
\affiliation{Physik-department, Technische Universit{\"a}t M{\"u}nchen, D-85748 Garching, Germany}
\affiliation{D{\'e}partement de physique nucl{\'e}aire et corpusculaire, Universit{\'e} de Gen{\`e}ve, CH-1211 Gen{\`e}ve, Switzerland}
\affiliation{Dept. of Physics and Astronomy, University of Gent, B-9000 Gent, Belgium}
\affiliation{Dept. of Physics and Astronomy, University of California, Irvine, CA 92697, USA}
\affiliation{Karlsruhe Institute of Technology, Institute for Astroparticle Physics, D-76021 Karlsruhe, Germany }
\affiliation{Karlsruhe Institute of Technology, Institute of Experimental Particle Physics, D-76021 Karlsruhe, Germany }
\affiliation{Dept. of Physics, Engineering Physics, and Astronomy, Queen's University, Kingston, ON K7L 3N6, Canada}
\affiliation{Dept. of Physics and Astronomy, University of Kansas, Lawrence, KS 66045, USA}
\affiliation{Department of Physics and Astronomy, UCLA, Los Angeles, CA 90095, USA}
\affiliation{Centre for Cosmology, Particle Physics and Phenomenology - CP3, Universit{\'e} catholique de Louvain, Louvain-la-Neuve, Belgium}
\affiliation{Department of Physics, Mercer University, Macon, GA 31207-0001, USA}
\affiliation{Dept. of Astronomy, University of Wisconsin{\textendash}Madison, Madison, WI 53706, USA}
\affiliation{Dept. of Physics and Wisconsin IceCube Particle Astrophysics Center, University of Wisconsin{\textendash}Madison, Madison, WI 53706, USA}
\affiliation{Institute of Physics, University of Mainz, Staudinger Weg 7, D-55099 Mainz, Germany}
\affiliation{Department of Physics, Marquette University, Milwaukee, WI, 53201, USA}
\affiliation{Institut f{\"u}r Kernphysik, Westf{\"a}lische Wilhelms-Universit{\"a}t M{\"u}nster, D-48149 M{\"u}nster, Germany}
\affiliation{Bartol Research Institute and Dept. of Physics and Astronomy, University of Delaware, Newark, DE 19716, USA}
\affiliation{Dept. of Physics, Yale University, New Haven, CT 06520, USA}
\affiliation{Dept. of Physics, University of Oxford, Parks Road, Oxford OX1 3PU, UK}
\affiliation{Dept. of Physics, Drexel University, 3141 Chestnut Street, Philadelphia, PA 19104, USA}
\affiliation{Physics Department, South Dakota School of Mines and Technology, Rapid City, SD 57701, USA}
\affiliation{Dept. of Physics, University of Wisconsin, River Falls, WI 54022, USA}
\affiliation{Dept. of Physics and Astronomy, University of Rochester, Rochester, NY 14627, USA}
\affiliation{Department of Physics and Astronomy, University of Utah, Salt Lake City, UT 84112, USA}
\affiliation{Oskar Klein Centre and Dept. of Physics, Stockholm University, SE-10691 Stockholm, Sweden}
\affiliation{Dept. of Physics and Astronomy, Stony Brook University, Stony Brook, NY 11794-3800, USA}
\affiliation{Dept. of Physics, Sungkyunkwan University, Suwon 16419, Korea}
\affiliation{Institute of Basic Science, Sungkyunkwan University, Suwon 16419, Korea}
\affiliation{Institute of Physics, Academia Sinica, Taipei, 11529, Taiwan}
\affiliation{Dept. of Physics and Astronomy, University of Alabama, Tuscaloosa, AL 35487, USA}
\affiliation{Dept. of Astronomy and Astrophysics, Pennsylvania State University, University Park, PA 16802, USA}
\affiliation{Dept. of Physics, Pennsylvania State University, University Park, PA 16802, USA}
\affiliation{Dept. of Physics and Astronomy, Uppsala University, Box 516, S-75120 Uppsala, Sweden}
\affiliation{Dept. of Physics, University of Wuppertal, D-42119 Wuppertal, Germany}
\affiliation{DESY, D-15738 Zeuthen, Germany}

\author[0000-0001-6141-4205]{R. Abbasi}
\affiliation{Department of Physics, Loyola University Chicago, Chicago, IL 60660, USA}

\author[0000-0001-8952-588X]{M. Ackermann}
\affiliation{DESY, D-15738 Zeuthen, Germany}

\author{J. Adams}
\affiliation{Dept. of Physics and Astronomy, University of Canterbury, Private Bag 4800, Christchurch, New Zealand}

\author[0000-0003-2252-9514]{J. A. Aguilar}
\affiliation{Universit{\'e} Libre de Bruxelles, Science Faculty CP230, B-1050 Brussels, Belgium}

\author[0000-0003-0709-5631]{M. Ahlers}
\affiliation{Niels Bohr Institute, University of Copenhagen, DK-2100 Copenhagen, Denmark}

\author{M. Ahrens}
\affiliation{Oskar Klein Centre and Dept. of Physics, Stockholm University, SE-10691 Stockholm, Sweden}

\author[0000-0002-9534-9189]{J.M. Alameddine}
\affiliation{Dept. of Physics, TU Dortmund University, D-44221 Dortmund, Germany}

\author{A. A. Alves Jr.}
\affiliation{Karlsruhe Institute of Technology, Institute for Astroparticle Physics, D-76021 Karlsruhe, Germany }

\author{N. M. Amin}
\affiliation{Bartol Research Institute and Dept. of Physics and Astronomy, University of Delaware, Newark, DE 19716, USA}

\author{K. Andeen}
\affiliation{Department of Physics, Marquette University, Milwaukee, WI, 53201, USA}

% ...

\collaboration{379}{IceCube Collaboration}

\affiliation{Science and Technology Institute, Universities Space Research Association, 320 Sparkman Drive, Huntsville, AL 35805, USA}
\affiliation{NASA Marshall Space Flight Center, Huntsville, AL 35812, USA}

\author{Adam Goldstein}
\affiliation{Science and Technology Institute, Universities Space Research Association, 320 Sparkman Drive, Huntsville, AL 35805, USA}

\author{Joshua Wood}
\affiliation{NASA Marshall Space Flight Center, Huntsville, AL 35812, USA}

\collaboration{2}{for the Fermi Gamma-ray Burst Monitor}

\date{\today}

\correspondingauthor{The IceCube Collaboration}
\email{analysis@icecube.wisc.edu}

\begin{abstract}

Gamma-ray bursts (GRBs) are considered as promising sources of ultra-high-energy cosmic rays (UHECRs) due to their large power output. Observing a neutrino flux from GRBs would offer evidence that GRBs are hadronic accelerators of UHECRs. Previous IceCube analyses, which primarily focused on neutrinos arriving in temporal coincidence with the prompt gamma rays, found no significant neutrino excess. The four analyses presented in this paper extend the region of interest to 14 days before and after the prompt phase, including generic extended time windows and targeted precursor searches. GRBs were selected between May 2011 and October 2018 to align with the data set of candidate muon-neutrino events observed by IceCube. No evidence of correlation between neutrino events and GRBs was found in these analyses. Limits are set to constrain the contribution of the cosmic GRB population to the diffuse astrophysical neutrino flux observed by IceCube. Prompt neutrino emission from GRBs is limited to $\lesssim$1\% of the observed diffuse neutrino flux, and emission on timescales up to $10^4$~s is constrained to 24\% of the total diffuse flux.

\end{abstract}


\section{Introduction}\label{sec:intro}

Gamma-ray bursts (GRBs) are short bursts of gamma radiation and are among the most energetic events in the Universe~\citep{2004IJMPA..19.2385Z,2006RPPh...69.2259M}. The primary burst of gamma rays, called the prompt emission, lasts for about $10^{-3}$~s to $10^3$~s. GRBs are broadly classified into two categories based on the duration of their prompt emission: short GRBs (for bursts shorter than 2~s) and long GRBs (for bursts longer than 2~s)~\citep{class_grbs_1,Class_GRBS_2,class_grbs_3}. Any particle emission observed prior to and after the prompt emission is referred to as precursor and afterglow emission, respectively. Short GRBs are generally observed to have a harder energy spectrum than long GRBs~\citep{hardness}. Although their exact emission mechanism is not well understood, the predominant model for GRB phenomenology includes the emission of a relativistic fireball triggered by the interaction of accreting matter onto a compact central object~\citep{GRB_overview,2006RPPh...69.2259M}. 
The recent observation of gravitational wave emission from a neutron star (NS) binary merger, GW170817, in coincidence with the short GRB 170817A~\citep{LIGO2017}, confirmed that short GRBs can be produced by mergers of compact objects. Long GRBs have been previously linked to the core collapse of super-massive stars by the observation of coincident supernovae~\citep{Hjorth_Bloom,Long_GRBs}. 

Central engines of GRBs drive a highly relativistic jet beamed into a narrow opening angle~\citep{2004IJMPA..19.2385Z,Beaming_GRB}. The jet fireball is hypothesized to be a plasma arising from a quasi-thermal equilibrium between radiation and $e^-e^+$ pairs. Multiple shells of plasma can be emitted, which propagate outward from the central engine into the interstellar region with a varying Lorentz factor~\citep{Paczynski,Goodman}. When two shells collide, a shock wave will develop that can accelerate charged particles to higher energies via first order Fermi acceleration~\citep{Fermi1,Fermi2}.  At a later stage, the relativistic outflow from the fireball will interact with the interstellar medium, leading to external shocks~\citep{2006RPPh...69.2259M}. In some models, protons and ions are accelerated at the sites of internal and external shocks to energies in excess of $10^{20}$~eV leading to emission of Ultra-High-Energy Cosmic Rays (UHECRs)~\citep{waxman_solo,Afterglow,Murase}.  UHECRs observed in coincidence with GRBs would offer direct evidence of this hadronic acceleration. However, cosmic rays get deflected by the intergalactic magnetic fields as they propagate through space. Thus they neither point back to their sources, nor reach us at the same time as the GRB gamma rays. Fortunately, neutrinos offer an alternative approach to identify the progenitors of cosmic rays. Fermi accelerated protons can interact with the gamma rays produced in the fireball and lead to the photo-meson production of pions, which can create an accompanying burst of neutrinos~\citep{Wax_Bahcall}. These interactions can take place through the following channels:

\begin{equation}
\begin{aligned}
p + \gamma &\rightarrow n + \pi^+\ ;\\
\pi^+ &\rightarrow \mu^+ + \nu_\mu\ ;\\
\mu^+ &\rightarrow e^+ + \Bar{\nu}_\mu + {\nu}_e\ .
\label{Nu_1}
\end{aligned}
\end{equation}

Assuming sufficient pion production, GRBs can collectively produce a diffuse neutrino flux observable at Earth above 0.1-1~PeV~\citep{Wax_Bahcall,1998PhRvD..58i3009G,Winter3,Winter2,Winter1,Winter4}. Neutrinos effectively only interact through the weak force, propagating through the Universe without deflection and thus point back to their sources. Detecting high-energy neutrinos correlated with GRBs would establish them as cosmic-ray acceleration sites.

The IceCube Neutrino Observatory is currently the most sensitive instrument for the detection of astrophysical neutrinos~\citep{Probing_parti_IceCube}. In 2013 the IceCube Collaboration first reported the discovery of an astrophysical neutrino flux~\citep{IceCubeHESEDiscovery}. This was later corroborated  by the discovery of a hard spectrum of muon events in the Northern Hemisphere~\citep{IceCubeDiffuseDiscovery}. While candidate neutrino sources have been identified by IceCube~\citep{TXS} and others~\citep{tde}, the origin of the diffuse astrophysical flux is not fully understood and may have several classes of progenitors.  Because searches for neutrinos in coincidence with brief, transient phenomena are nearly background free,  IceCube has pursued several different types of analyses designed to search for neutrino correlations with the prompt phase of GRB observations \citep{IceCubeGRB2012,IceCubeGRB2015,IceCubeGRB2016,IceCubeGRB2017}, but found no associations. These results are consistent with non-detections in analyses performed by AMANDA~\citep{AMANDA_2,AMANDA_1} and ANTARES~\citep{antares2013,ANTARES2020,501}. The most recent IceCube results~\citep{IceCubeGRB2017,502} have put constraints on the single-zone fireball models of GRB neutrino and UHECR production during the prompt phase. IceCube has also performed dedicated searches for neutrinos coincident with gravitational wave candidates, such as GW170817, detected by LIGO and Virgo~\citep{2016PhRvD..93l2010A,2017PhRvD..96b2005A,2017ApJ...850L..35A,IceCubeGW1,IceCubeGW2}. To date, no correlation has been found on $10^3$~s time scales, but additional studies are ongoing \citep{SwiftFollowUp}.

Recent observations by imaging air Cherenkov telescopes (IACTs) have shown that TeV particles can be produced during the afterglow phase, up to several days after the prompt emission \citep{194,328,TeVGRBAfterglow}. Additionally, it has been shown that $\sim$10\% of GRBs have an observed gamma-ray precursor that precedes the main burst by a few tens of seconds, but in extreme cases, up to 10 minutes (e.g. \cite{108,109,Precursors}). The complementary studies presented in this paper extend the search for neutrino correlations to precursor time windows, as well as extended precursor and afterglow time windows of up to $-14$ to $+14$ days around GRB gamma-ray triggers.  The low background rates allow for highly sensitive searches on the scale of days to weeks.  All the analyses use 7.16 years of IceCube muon neutrino candidate events. Section 2 describes the IceCube detector and the event selection for the neutrino dataset used in the four analyses. In Section 3, we describe the catalog of GRBs that was used in our analyses, as well as the different selection cuts on the GRB sample that were considered for the respective analyses. In Section 4, we describe the methods used for evaluating the statistical significance of results, and in Section 5 we describe the analysis approach and results for each study. Our limits and interpretation on neutrino emissions from cosmic GRB populations are presented in Section 6. Section 7 then provides the concluding remarks and outlook. 

\section{Detector \& Event Selection}

The IceCube Neutrino Observatory is a cubic-kilometer-scale Cherenkov detector buried deep in the South Pole ice. IceCube consists of 5,160 digital optical modules (DOMs) arranged in an array of 86 strings deployed 1,450 to 2,450 meters below the ice surface~\citep{IceCubeDetector}.  The strings are arranged in a hexagonal grid with 125~m spacing between adjacent strings and with each string containing 60~DOMs, spaced 17~m apart along the string. Each DOM houses a downward facing photo-multiplier tube (PMT) inside a spherical transparent glass capsule. The DOMs are designed to be sensitive to the Cherenkov radiation produced by the secondary particles that result from interactions of neutrinos with the ice. Cherenkov radiation is produced when a charged particle travels faster than the speed of light in a dielectric medium, resulting in conical emission of photons along the path of the charged particle. DOMs record the waveforms of Cherenkov photons, from which the number of photo-electrons and their arrival time can be extracted \citep{IceCubeDetector}. This information is combined from all DOMs to reconstruct the Cherenkov light cones and infer the energy and direction of the particles that produced them. 

IceCube is sensitive to all three flavors of (anti-)neutrinos; however, the data set used in these analyses is optimized to select charged-current interactions from muon (anti-)neutrinos, as they offer the best pointing resolution. These interactions result in the production of a muon that will propagate through the detector in a straight line, depositing light along its track. The  typical angular resolution for these tracks is $\lesssim 1$\textdegree~for muons with energies $\gtrsim1$~TeV. 

The sample used for this paper is the IceCube gamma-ray follow-up (GFU) data consisting of well-reconstructed muon tracks collected from 2011-05-13 through 2018-10-14~\citep{GFUref}. The vast majority of events that trigger the IceCube detector are not astrophysical neutrinos, but muons produced in cosmic-ray air showers. In the Southern hemisphere, atmospheric muons are the dominant background and are observed at a rate of 2.7 kHz~\citep{IceCubeDetector}. Since only neutrinos can propagate through the Earth without being absorbed, this background vanishes in the Northern hemisphere, where atmospheric neutrinos dominate the background. A selection with different data quality cuts for the Northern and Southern hemisphere is therefore used, which reduces these backgrounds to 6.6~mHz integrated over the full sky~\citep{GFUref}. A detailed account of this event selection is given in \citep{GFUref}. 

\section{GRB Catalog}

Space-based gamma-ray observatories, such as \emph{Swift} and \emph{Fermi}, as well as a variety of ground-based observatories, continuously monitor the sky for high-energy gamma-ray activity~\citep{Swift_mission,Fermi_GBM_paper}. These observatories provide hundreds of GRB measurements per year, which we used to construct a GRB catalogue to enable our coincidence study. GRBweb \citep{GRBweb} is an IceCube project designed to combine the observational data from all major GRB observatories into a single database. Input data to GRBweb primarily originates from online GRB catalogs, such as those by \emph{Fermi} \citep{GBMGRBcat,LATGRBcat}, \emph{Swift}~\citep{SwiftGRBcat}, and IPN~\citep{IPNGRBcat}, and from the automated parsing of GCN circulars~\citep{GCN}. An all-inclusive GRB catalog is thus constructed. Internally, GRBweb makes use of a set of automated Python scripts that process and save the data into a SQL database. GRBweb currently contains over 7,500 GRBs and is updated on a weekly basis.

GRBweb uses a set of predefined conditions\footnote{For a detailed description of all variables and selection criteria, see \url{https://icecube.wisc.edu/~grbweb_public/Variables.html}.} to determine which data will be used if the burst was observed by multiple detectors. For instance, the burst direction is set equal to the localization with the smallest angular uncertainty. Burst times, $T_0$, correspond to the earliest time at which gamma-ray activity was reported. For the GRB duration, two variables are considered: the conventional $T_{90}$ and new, composite variable called the $T_{100}$.  The $T_{100}$ is defined as the time difference between the last and earliest reported time of gamma-ray activity.

Each analysis selected a subset of GRBs between 2011 and 2018. The selection was motivated by the effect of localization uncertainties on the sensitivity of the analysis, as well as requirements on timing or precursor information.  The differences between the analyses is summarized in Table \ref{tab:GRBs} and explored in detail in Section \ref{overview}. As the data set ends in October 2018, the only IACT-detected burst in our sample is GRB~180720B \citep{328}, which is discussed in Section \ref{sec:results}.

\begin{table}\centering
    \caption{The maximum uncertainty on the localization, the number of GRBs used in each analysis, and the number of GRBs localized only by~\emph{Fermi}/GBM.}
    \begin{tabular}{|*{1}{C{3.5cm}} *{1}{C{5.0cm}} *{1}{C{2.0cm}} *{1}{C{5cm}}|}
        \hline
        \textbf{Analysis} & \textbf{Max Localization Uncertainty} & \textbf{\# GRBs} & \textbf{\# Localized only by GBM} \\
        \hline
        Extended TW & no max & 2091 &  1236 \\
        Precursor/Afterglow & 0.2$^\circ$ & 733 & 0 \\
        GBM Precursor & no max & 133 & 100 \\
        Stacked Precursor & 1.5$^\circ$ & 872 & 0 \\
        \hline
    \end{tabular}
    \label{tab:GRBs}
\end{table}

\section{Statistical Method}\label{analysis_methods}
Each of the analyses makes use of an unbinned likelihood ratio method to quantify the potential correlation between GRB observations and IceCube events, using a blind analysis technique. In this section, we present details on the likelihood ratio method used to determine the \textit{p}-value of individual GRBs, as well as the binomial test used to assess the group of \textit{p}-values from all GRBs.

\subsection{Likelihood Ratio \& Test Statistic}\label{sec:llh}

An unbinned likelihood ratio method \citep{LLH1,LLH2,IceCubeGRB2017} is combined with frequentist statistics to assign a probability that a subset of neutrino candidate events is consistent with background. For a sample of $N$ candidate neutrino events with characteristics $x_i$, the likelihood can be written as
\begin{equation}
    \mathcal{L}(n_s|n_b,{x_i}) = P_N
    \prod_{i=1}^N \left[p_s S(x_i)+p_b B(x_i) \right]\ ,
    \label{eq:likelihood}
\end{equation}
where $p_s=n_s/(n_s+n_b)$, $p_b=n_b/(n_s+n_b)$, and $P_N$ is the Poisson probability to observe $N$ events, assuming $n_s$ signal events and $n_b$ expected background events,
\begin{equation}
    P_N = \frac{(n_s+n_b)^N}{N!}e^{-(n_s+n_b)}\ .
\end{equation}
$S$ and $B$ denote the probability distribution functions (PDFs) describing the spatial and energy distribution of signal events and background events, respectively.
For the signal energy PDF, an $E^{-\gamma}$ power-law spectrum is assumed, where the spectral index $\gamma$ is either a fit parameter or is fixed to $\gamma=2$, depending on the specific analysis (details in Section~\ref{overview}). The signal space PDF, shown in Eq.~\eqref{sig_space_pdf}, uses a 2D Gaussian to test the compatibility of the neutrino candidate's reconstructed position, $\vec{x}_\nu$, with the source position, $\vec{x}_{GRB}$,
\begin{equation}\label{sig_space_pdf}
    S_{\text{sp}}(\vec{x}_\nu, \sigma | \vec{x}_{GRB}) = \frac{1}{2 \pi \sigma^2} \exp \left( -\frac{|\vec{x}_\nu - \vec{x}_{GRB}|^2}{2 \sigma^2} \right)\ ,
\end{equation}
where $\sigma^2$ is the quadratic sum of the uncertainty on the reconstructed neutrino and GRB direction. Since the contribution of signal events to the total data set is expected to be extremely small, the data can be used to construct the background PDFs.  In particular, the background energy PDF is assumed to follow the energy distribution of data. Due to the detector geometry, the approximation of azimuthal symmetry can be used to describe the background space PDF solely as a function of zenith. Any neutrino emission that occurs within a given analyzed time window is fitted as constant emission during the time window. The specific time windows used in each analysis are described in Section~\ref{overview}.

The likelihood Eq.~\eqref{eq:likelihood} is evaluated for different values of $n_s$ using the PDFs described above, with $\hat{n}_s$ designating the value of $n_s$ that maximizes the likelihood.  A likelihood ratio with respect to the null hypothesis $L(n_s=0)$ is then used to obtain the following test statistic (TS):
\begin{equation}\label{eq:TS_formal}
    \text{TS} = 2\cdot \ln \left[\frac{L(\hat{n}_s)}{L(n_s=0)} \right] = -2\hat{n}_s + 2\sum^N_{i=1} \ln \left[\frac{\hat{n}_s S(x_i)}{ n_b B(x_i)} +1 \right]\ .
\end{equation}
High-energy events in temporal and close spatial coincidence with a GRB will result in a large TS value. The $p$-value of an observation is determined by comparing the observed test statistic to a test statistic distribution created from scrambled data (see Section~\ref{sub:Sensitivity}).

Two of the four analyses use localizations provided by \emph{Fermi}-GBM for some GRBs and therefore have an additional step to determine the test statistic. Starting in early 2018, the Fermi-GBM collaboration began releasing a HEALPix skymap~\citep{433} with the localization probability as a function of sky position for each GRB localized by GBM~\citep{Gold_GBM}. Maps prior to 2018 were processed in a similar way using the GBM Data Tools, but the metadata in the files have not been fully qualified and the files have not yet been uploaded to the final HEASARC archive, therefore we use the preliminary files from~\cite{Goldstein2022}. These maps contain a per-pixel probability that a given GRB originates from that direction. An all-sky scan of neutrino data is performed, in which TS$_\text{orignial}$ (equal to TS from Eq. \eqref{eq:TS_formal}) is calculated at every pixel of the skymap and then penalized by the probability, $P_\mathrm{GBM}$, of that pixel
\begin{equation}\label{eq:TS_prior}
    \text{TS}_\text{final} = \text{TS}_\text{original} + 2\times \left[\ln(P_\mathrm{GBM})\ - \ln(P_\mathrm{GBM,max})\right]\ ,
\end{equation}
where $P_\mathrm{GBM,max}$ is the maximum probability on the entire skymap.  The position of the GRB on the sky and the number of signal events are thus both fitted to find the combination which maximizes $\mathrm{TS_{final}}$.

\subsection{Stacked Likelihood Analysis}\label{sec:stacked_llh}
Equation \eqref{eq:likelihood}, which describes the likelihood for a single well-localized GRB,  can be easily modified to describe the likelihood of $N_{GRB}$ well-localized GRBs.  When considering multiple sources, $n_s$ corresponds to the total number of signal events summed over all GRBs. Each GRB is assumed to have the same time-integrated neutrino flux at Earth. The expected number of neutrinos from each GRB will therefore be proportional to the effective area at the declination of the burst, $\mathcal{A}_{eff}(\delta)$, which is calculated assuming an $E^{-2}$ neutrino energy spectrum.
The signal PDF is then replaced by the sum of the $N_{GRB}$ signal PDFs, weighted by their relative contribution to the number of signal events
\begin{equation}
	S=\frac{\sum_{j=1}^{N_{GRB}} \mathcal{A}_{eff}(\delta_j)\cdot S_j }{ \sum_{j=1}^{N_{GRB}} \mathcal{A}_{eff}(\delta_j) }\ ,
	\label{eq:DetectorWeight}
\end{equation}
where $S_j$ and $\delta_j$ are the signal PDF and declination of the $j$'th burst, respectively. This method, known as a stacked likelihood analysis, provides a single measure of the total neutrino emission from a set of GRBs.

\subsection{Cumulative Binomial Test}\label{binomial_test}

When not performing a stacked search, analyzing a selection of $N$ GRBs will result in a $p$-value for each individual burst. A trial-correction method is thus needed to determine if one or more of the obtained $p$-values provides a statistically significant result. Arranging the $p$-values from smallest to largest, their values are denoted as $p_1,p_2, ..., p_N$. Under the null hypothesis of no neutrino emission, the correlations of neutrino events with GRBs will only occur randomly, and these $N$ $p$-values are expected to follow a uniform distribution between 0 and 1. The probability that $k$ or more $p$-values are smaller than or equal to $p_k$ is thus given by the binomial probability:

\begin{equation}
    P(k) \equiv P(n\geq k|N,p_k) = \sum_{m=k}^N \frac{N!}{(N-m)!m!} p_k^m (1-p_k)^{N-m}\ .
\label{eq:binomial}
\end{equation}

\begin{figure}[t]
    \centering
    \includegraphics[width=0.49\linewidth]{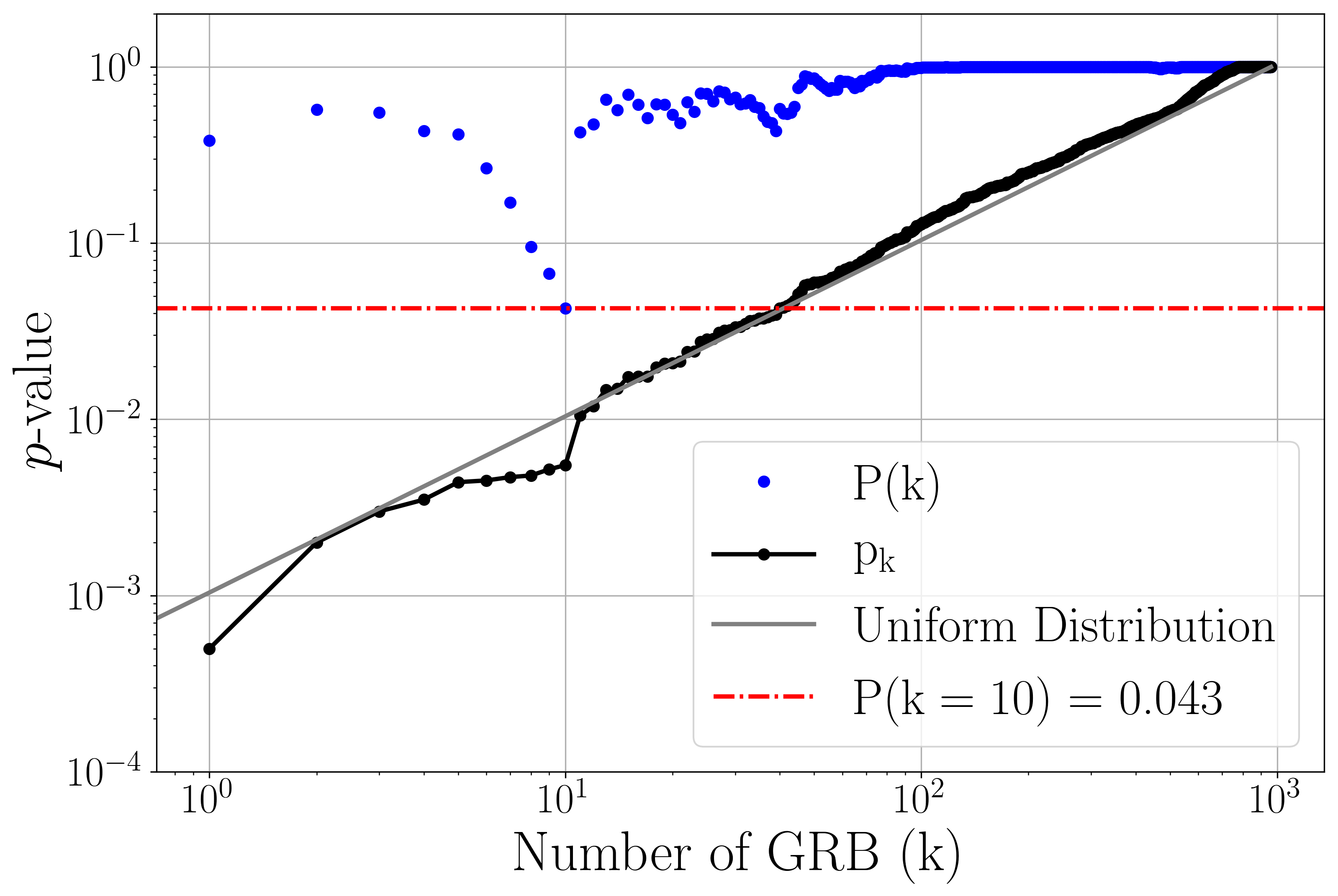}
    \includegraphics[width=0.49\linewidth]{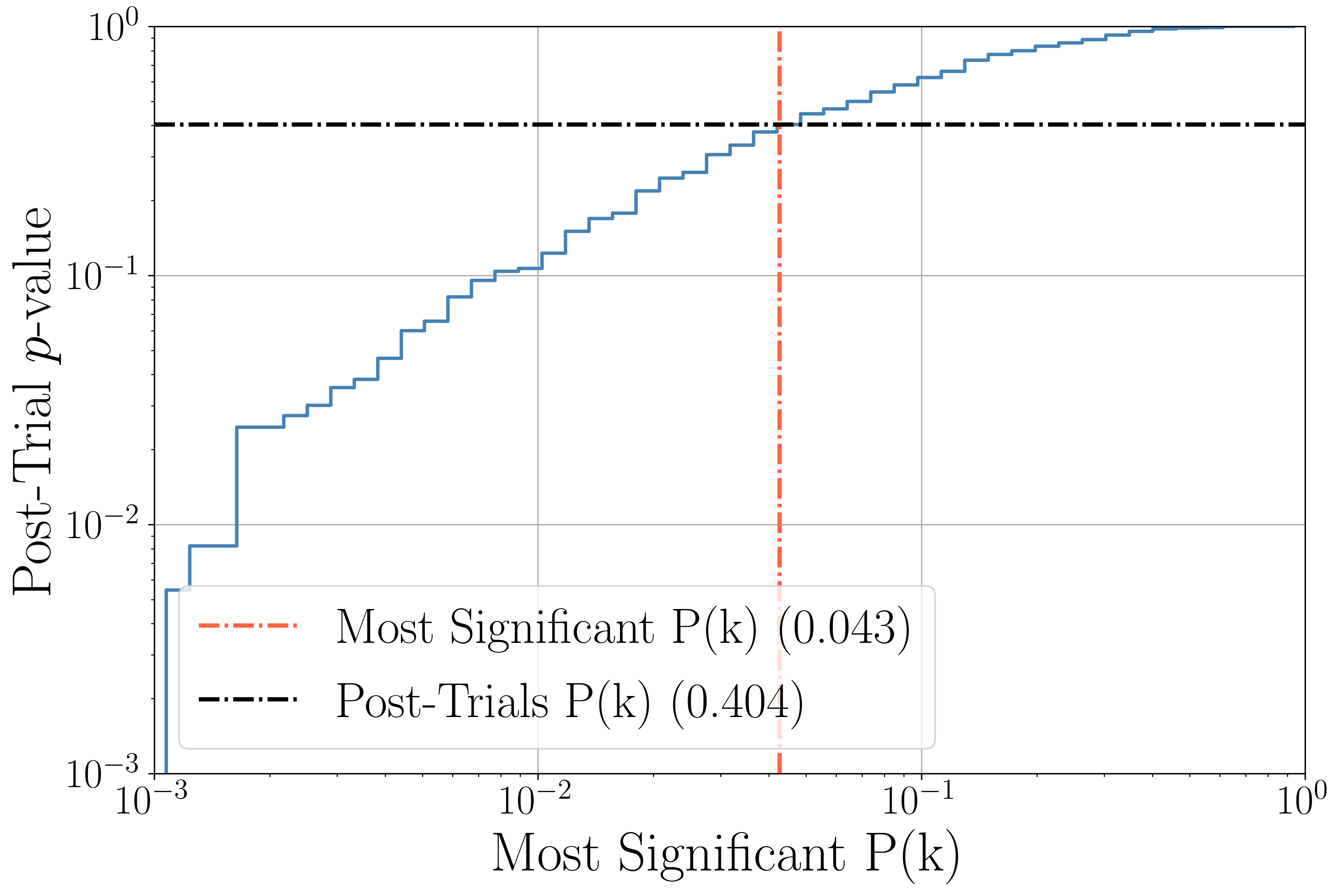}
    \caption{Left: An example of the binomial test based on simulation.  The black dots
    show the ordered list of $p_k$ values constructed from the pre-trial \textit{p}-values $p_{pre}$ of every analysed GRB.  Each blue dot shows the corresponding binomial \textit{p}-value $P(k)$ defined in Eq.~\eqref{eq:binomial}.  In this example, $k=10$ (i.e.\ the ten GRBs with the smallest $p_{pre}$ values) was found to be most significant subset, with $P(k)=0.043$ (red line).  Note that the accumulation of $p=1$ at large $k$ is typical for low-background searches (GRBs with $\hat{n}_s=0$ are assigned $p=1$).  Right: The trial correction process is shown, with the most significant $P(k)$ from the left plot corresponding to a post-trial \textit{p}-value of 0.4.  The cumulative distribution in this plot is created by performing the analysis on  scrambled data sets.}
    \label{fig:binom_test_example}
\end{figure}

\noindent Evaluating Eq. \eqref{eq:binomial} over all potential values of $k$, the smallest $P(k)$ is selected. An empirical trial-correction factor is then applied to account for the fact that $N$ potential $p$-values were scanned. This trial correction is found by determining the fraction of background-only realizations that produce a more significant result. A visualization of this trial-correction procedure is shown in Figure \ref{fig:binom_test_example}. This procedure also ensures that the rare overlap of a single neutrino contributing to two analyzed GRBs (thus creating a correlation between two $p$-values) is also accounted for in the final post-trial p-value.
On the other hand, given that the successive probabilities $P(k)$ are strongly correlated with one another, the overall trial correction factor is modest in the end.

\subsection{Sensitivity and Upper Limits}\label{sub:Sensitivity}
To test the neutrino flux to which the analyses are sensitive, simulated
muon neutrino and muon anti-neutrino events based on full detector Monte Carlo (MC) can be injected into the data sample. The energy spectrum of the MC-generated neutrinos can be fixed to $E^{-2}$ or other spectra as desired. To describe the background, we use data with event times that are randomly selected from a uniform distribution over the livetime of the dataset while accounting for the detector downtime. Keeping the event coordinates fixed in the detector frame, scrambling background event times correspondingly randomizes the right ascension. Such a pseudo-dataset is called \emph{scrambled data}. Signal can then be ``injected'' by adding the signal-like events to it. In each realization of scrambled data plus signal, the number of injected MC events is drawn from a Poisson distribution with a fixed mean $\mu$. By varying the value of $\mu$, the threshold can be determined at which 90\% of all realizations result in a TS value that is larger than the median of the background TS distribution of the analysis. We define this threshold, and the neutrino flux to which it corresponds, as the sensitivity of the analysis. Once the analysis has been performed, upper limits can be calculated in a similar way, by comparing to the observed TS in data rather than the median TS of the background-only realizations.

\section{Analysis}\label{overview}
IceCube has previously reported limits on neutrino production during the prompt phase of GRB observations and found no association, yielding limits that have constrained several leading models of GRBs \citep{IceCubeGRB2017}.  However, recent observations of gamma-ray emission outside of the prompt phase motivate a more comprehensive search. The four analyses described below allow for the possibility of neutrino emission outside of the prompt phase. As with previous GRB searches, all analyses are primarily based on the spatial and temporal correlation between neutrino events and GRB observations, while they differ in their specific assumptions about the neutrino emission. The first two analyses are the most general, with the ``Extended TW'' analysis opening the observation window to up to one day prior and two weeks following the prompt emission. The ``Precursor/Afterglow'' analysis focuses on a smaller sample of well-localized GRBs in our catalogue, while treating precursor emission separately from emission during and after the prompt phase.  The final two analyses focus specifically on the precursor phase, with the ``GBM Precursor'' analysis examining GRBs for which Fermi-GBM detected gamma-ray emission prior to the prompt phase, while the ``Stacked Precursor" analysis examines well-localized bursts for which no precursor emission was observed.  Together, these analyses provide both a comprehensive and a model-independent approach to the search for neutrino production in GRBs.  Each analysis and its results are described in more detail below. Limits and interpretation of the results are then presented in Section \ref{sec:results}.

\subsection{Extended TW}
This analysis searches for neutrinos in a range of time windows (TW) and uses the largest GRB catalog of the four analyses. All GRBs with known duration were included, given they were observed during the detector livetime. 163 GRBs do not have a reported $T_{100}$ in GRBweb at the time of writing, and so these GRBs were excluded from the selection.  The final number of GRBs in this analysis is 2,091.

\begin{figure}
    \centering
    \includegraphics[width=0.9\linewidth]{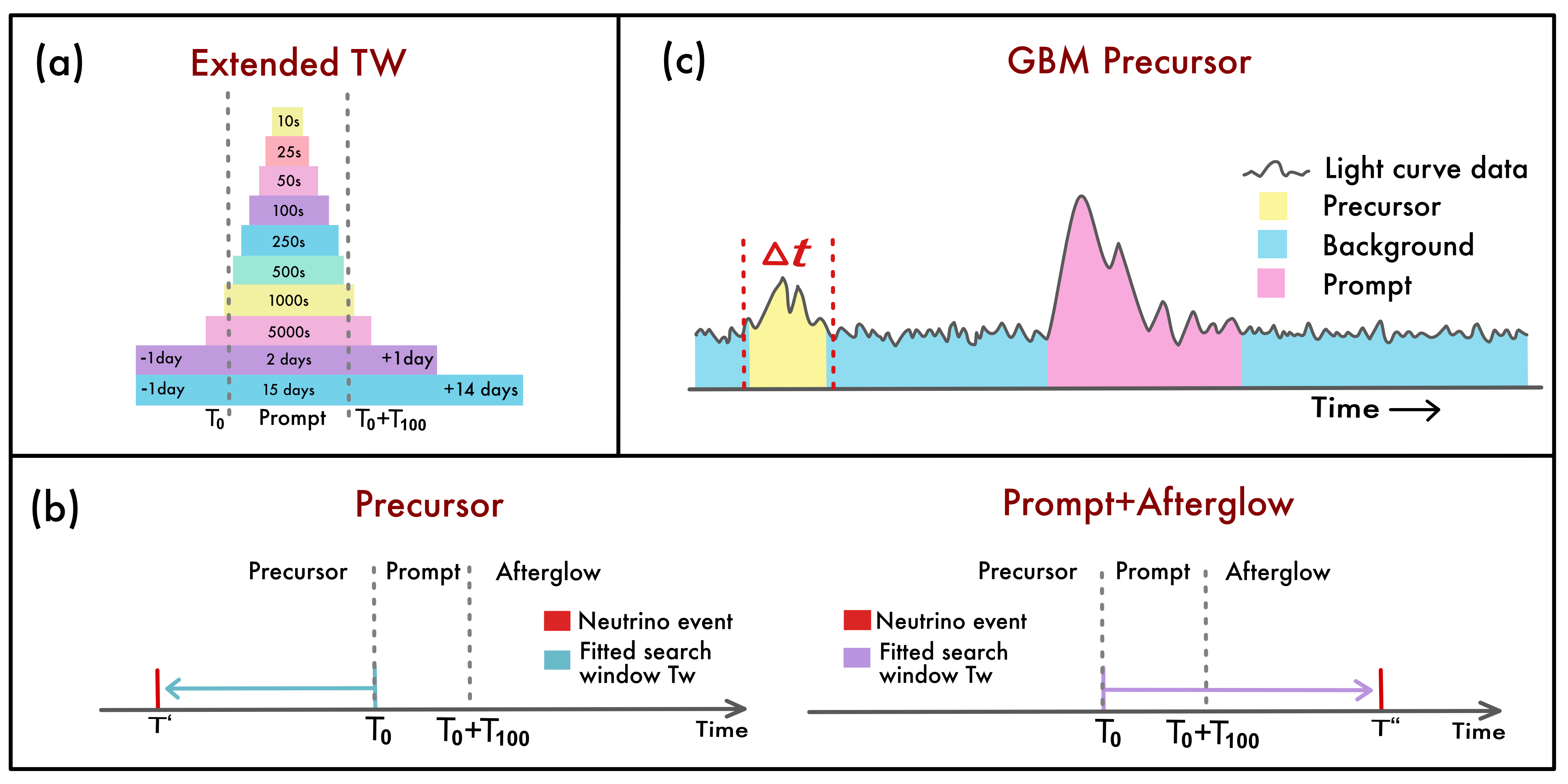}
    \caption{(a): The time windows used in the ``Extended TW'' analysis.  These 10 time windows are evaluated for all GRBs, regardless of their duration. The first 9 time windows are centered on the $T_{100}$, while the longest time window is asymmetric around the $T_{100}$. The $T_{100}$ of a given GRB is the duration between the earliest and latest measured time of gamma-ray activity. The shortest time window that completely contains the $T_{100}$ is used to determine the prompt limit. In the example shown above, the prompt time window is $10^3$ seconds. (b) A schematic representation of how the best-fit time windows are obtained for the precursor (left) and the prompt+afterglow searches (right) in the ``Precursor/Afterglow'' analysis. (c) Illustration of a gamma-ray light curve in which the GRB prompt phase is preceded by a precursor. The time range of the precursor, extended by 2~s on either side, is used to search for coincident neutrinos in the ``GBM Precursor'' analysis.}
    \label{extended_tw_cartoon}
\end{figure}

The $p$-values are calculated for these 2,091 GRBs in 10 pre-determined time windows and with a fixed $E^{-2}$ neutrino energy spectrum in the signal hypothesis. The first 9 time windows range from 10 seconds to 2 days, centered on the $T_{100}$ of the GRB, and the final time window is asymmetric with a 1 day precursor and 14 day afterglow time window (see Figure \ref{extended_tw_cartoon}a). The shortest pre-defined time window that fully envelopes the $T_{100}$ interval will be used to describe the prompt emission. The seven shortest time windows range up to $10^3$ seconds, which includes most GRB prompt phases.  The 10 second time window is sufficiently small to study the short GRBs, because the neutrino background is effectively zero at this timescale. For the longer time windows the background becomes non-negligible, which is what motivates the decision to search up to $\sim$ two weeks but not longer. For a well-localized GRB ($\sigma\lesssim 1^\circ$), the 15-day time window has an expected background on the order of one neutrino candidate event. For poorly-localized GRBs this longer window is less sensitive, which is why the emphasis was placed on the afterglow region where higher-energy neutrinos are predicted \citep{asano_murase_2015}. The test statistic in each time window is calculated using Eq.~\eqref{eq:TS_prior}. From those 10 $p$-values, the most significant one is selected to represent the GRB.

Since each GRB is studied in 10 time windows, a correction is required to compensate for the look-elsewhere effect.  Because the time windows are correlated, an effective trial correction is used. For every given GRB, the smallest $p$-value of the 10 time windows is selected. Background scrambles are then used to determine the probability of obtaining a value that is smaller than or equal to this result.  Next, the \textit{p}-values that have been corrected for searching 10 time windows are evaluated in a binomial test (Section \ref{binomial_test}) to search for evidence of a sub-set of GRBs with significant neutrino emission.

Four binomial tests are evaluated, where the total GRB sample has been divided into four sub-populations by duration and hemisphere. Following~\cite{vonK2020}, GRBs are separated into short and long classes according to whether the measured T90 is less than or greater than 2~s, respectively. This is intended to account for the different progenitor classes of merger events~\citep{LIGO2017} and core collapse supernovae~\citep{Hjorth_Bloom}, albeit in a simplistic way since the two classes are known to overlap. This overlap explains why GRB~170817A falls into the long GRB class with a T90 of 2.048~s despite being a known short GRB from a binary neutron star merger. In this case, its misclassification does not notably affect the results as GRB~170817A would not have appeared in the top 5 short GRBs from the Southern Hemisphere (listed in Appendix~\ref{appendix_A_tables}) even if it were included in that class. Future studies will explore grouping methods which account for the fact that GRB~170817A is a short GRB. The split by hemisphere allows for the increased sensitivity of the analysis to GRBs in the Northern sky.

The final p-value for each binomial test is determined by comparing the result found for data with the results found for scrambled data sets. The final p-values are summarized in Table \ref{liz_binomial_results} and are consistent with background.  The most significant GRB (pre-trial) from each sub-population is listed in Table \ref{liz_top_grbs}. All GRBs with a $p$-value less than 1\% are listed in Appendix \ref{appendix_A_tables}.

\begin{table}
    \centering
    \caption{Post-trial \textit{p}-value of the binomial test for the ``Extended TW'' study of 2,091 GRBs. The binomial test was run on four subsets of GRBs split by hemisphere and prompt gamma-ray duration.  The number of GRBs in each sub-population is indicated in parentheses.}
    \begin{tabular}{|*{4}{C{4cm}}|}
        \hline
        Northern Long (960) & Northern Short (183) & Southern Long (814) & Southern Short (134) \\
        \hline
        0.038 & 0.799 & 0.898 & 0.849 \\
        \hline
    \end{tabular}
    \label{liz_binomial_results}
\end{table}

\begin{table}
    \centering
    \caption{The top GRB result found in each sub-population in the ``Extended TW'' study of 2,091 GRBs.  The number of GRBs in the sub-population is indicated in parentheses.  The column titled $p_{pre}$ gives the p-value for the GRB, without a correction for the size of the sub-population.  This $p_{pre}$ p-value has been corrected for searching 10 time windows.  The column $p_{post}$ provides the corresponding p-value including the look-elsewhere correction for both the population size and for searching ten time windows.}
    \begin{tabular}{|*{2}{C{3.5cm}} *{1}{C{5.0cm}} *{2}{C{2.0cm}}|}
        \hline
        GRB Name & Sub-Population & Most Significant Time Window & \textit{p}$_{\mathrm{pre}}$ & \ \textit{p}$_{\mathrm{post}}$ \\
        \hline
        GRB 140607A & Northern Long (960) & $\pm$1 Day & 6.0e-04 & 4.4e-01 \\
        GRB 140807500 & Northern Short (183) & 100 Seconds & 4.8e-03 & 5.9e-01 \\
        GRB 150202A & Southern Long (814) & $\pm$1 Day & 5.0e-04 & 3.3e-01 \\
        GRB 140511095 & Southern Short (134) & $\pm$1 Day & 9.2e-03 & 7.1e-01 \\
        \hline
    \end{tabular}
    \label{liz_top_grbs}
\end{table}

\subsection{Precursor/Afterglow}

In the ``Precursor/Afterglow'' analysis, the time window for the signal region is treated as a parameter which can be estimated from the data. The time window can be fitted from a minimum duration of 0.5~s up to a maximum of 2 weeks, long enough to cover a wide range of possible neutrino emission time scales. Only GRBs with a positional angular uncertainty of less than 0.2\textdegree\ are considered in this analysis (thus they can be approximated as point sources), which keeps the background low within the full time window range. GRBs within the first 14 days and the last 14 days in the GFU data are further removed for this analysis. This results in a selection of 733 GRBs. This includes 53 GRBs that did not have a reported $T_{100}$ in GRBweb and were not included in the Extended TW analysis. Two separate searches are performed for every GRB, denoted as the ``precursor'' search and ``prompt+afterglow'' search, each with fit parameters corresponding to the number of signal events ($n_s$), the spectral index ($\gamma$), and the width of the emission time window ($T_w$). The only difference between the two searches is how the time window $T_w$ is defined with respect to the start of the prompt phase $T_{0}$ (see Figure~\ref{extended_tw_cartoon}b):

\begin{itemize}
\item For precursor searches, the time window extends backwards from $T_0$ to the time ($T_0 - T_w$) before the start of the prompt phase.

\item For prompt+afterglow searches, the time window extends forwards from $T_0$ to the time ($T_0+T_w$) after the start of the prompt phase.
\end{itemize}

\begin{figure}
    \centering
    \includegraphics[width=0.60\linewidth]{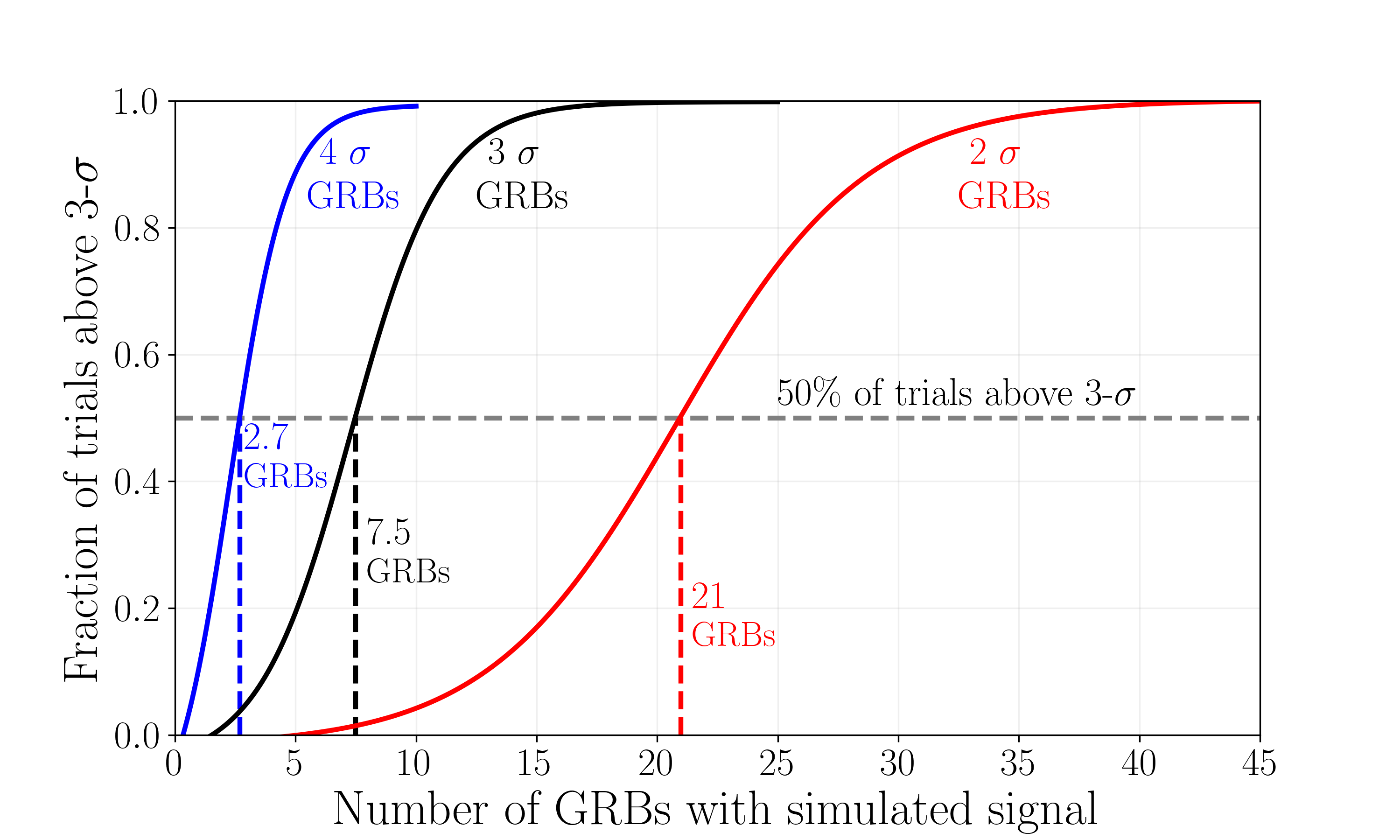}
    \caption{Post-trial 3$\sigma$ discovery potential for the Precursor/Afterglow search. The $x$-axis shows the number of injected GRBs of equal strength. GRBs with 2$\sigma$ (red), 3$\sigma$ (black), and 4$\sigma$ (blue) significance have been injected. The $y$-axis shows the fraction of $10^4$ simulated binomial tests that have a final \textit{p}-value of 3$\sigma$ significance. The x-axis shows the Poisson mean of injections. }
    \label{3sig_plot}
\end{figure}

For every GRB, the analysis returns the best-fit parameters for the respective search and the  \textit{p}-value. This results in two lists of 733 \textit{p}-values, one list for each search.  A binomial test (Section \ref{binomial_test}) is performed on each list to search for a subset of GRBs with significant neutrino emission. The final \textit{p}-value after each binomial test is determined by comparing the result found for data with the results found for scrambled data sets.
Figure~\ref{3sig_plot} demonstrates different numbers of individual GRBs with 2$\sigma$, 3$\sigma$ and 4$\sigma$ \textit{p}-values that can typically result in a final post-trial \textit{p}-value of 3$\sigma$ significance.

The final post-trial \textit{p}-values of the binomial tests are 0.495 for the precursor search and 0.486 for the prompt+afterglow search. The results are comparable with the median background expectation (i.e.\ \textit{p}-value of 0.5). The results for the top 20 GRBs in each search are summarised in Appendix~\ref{AppendixB}.

\subsection{GBM Precursor} 

The two analyses above searched for neutrino correlations with GRBs in an agnostic manner. It is possible to use additional gamma-ray observations to perform more sensitive dedicated analyses.
A previous study by~\cite{Precursors} analyzed the light curves of all \emph{Fermi}-GBM bursts up to 2020, leading to a sample of 217 GRBs that exhibit signs of gamma-ray precursors. Of those 217 bursts, there are 133 GRBs that overlap with the IceCube data set examined here. A dedicated search is performed to investigate potential neutrino production coincident with the emission phase of the observed gamma-ray precursors using these 133 GRBs. The time windows of the analysis are set to those of the identified gamma-ray precursors, extended by 2~s on either side to obtain a restrictive yet conservative range. Summed over all 133 GRBs, a total time window of $3.3\cdot 10^3$~s is examined. The signal hypothesis in the likelihood uses a fixed $E^{-2}$ neutrino energy spectrum.
Out of 133 bursts, 100 GRBs were localized solely by \emph{Fermi}-GBM. For those 100 bursts, a TS is defined following Eq. \eqref{eq:TS_prior}. Otherwise, the TS is based on Eq. \eqref{eq:TS_formal}.

Performing the analysis on all GRBs results in 133 individual $p$-values. To trial correct the result, a procedure similar to that of the binomial correction is applied. However, instead of considering the $k$th most significant $p$-value, the product of the $k$ most significant $p$-values is considered. Comparing this result to the distribution of the same statistic (product of the $k$-most significant \textit{p}-values) found in scrambled data sets yields the final \textit{p}-value for the analysis.

As this analysis only targets GRBs with observed gamma-ray precursors, it has the smallest background of all four analyses. 
The neutrino flux to which the analysis is sensitive (Sec.~\ref{sub:Sensitivity}) corresponds to 
only 1.7 neutrino events on average. Considering events within a relative combined neutrino and GRB angular uncertainty of $5\sigma$, no neutrino events were observed in temporal coincidence with any of the 133 precursors. A final $p$-value of 1 was thus obtained.

\subsection{Stacked Precursor}

One of the results by \cite{Precursors} was that almost all ($>$95\%) gamma-ray precursors were found to occur within 250~s of the prompt emission. A fourth analysis was therefore performed to enable a larger systematic search for precursor neutrinos, not limited to \emph{Fermi}-GBM bursts for which a gamma-ray precursor could be resolved. For all well-localized\footnote{A less restrictive cut of 1.5$^\circ$ is applied on the localization compared to the ``Precursor/Afterglow'' analysis, as the significantly smaller time window and reduced number of fit parameters lead to a smaller effective background.} GRBs within the IceCube GFU data period, corresponding to 872 bursts, a generic time window of 250~s prior to the $T_0$ time was searched for excess neutrinos. All 33 well-localized GRBs from the GBM Precursor analysis are included in this search. For the set of 872 GRBs, a single stacked test statistic is constructed according to Eq. \eqref{eq:TS_formal} and assuming an $E^{-2}$ signal spectrum. Due to the stacking procedure, this analysis automatically leads to a single $p$-value, not requiring a trial correction procedure.

When the stacked precursor analysis is applied to these 872 GRBs, no excess of neutrino events is found ($n_s=0$).  Overall, five low-energy events within the given time window are in loose spatial coincidence with the examined GRBs, listed in Tab. \ref{tab:res_Prec_stack}. Given the length of the time window and the number of bursts, these coincident events are fully consistent with the background expectation. Similar to the GBM precursor analysis, a final $p$-value of 1 is thus obtained.

\begin{table}[H]
\caption{Properties of the neutrinos arriving in coincidence with GRBs in the Precursor Stacking search. For each neutrino, the angular separation from the GRB, angular uncertainty of the neutrino direction, a proxy for the event energy and the arrival time before the GRB is shown.}
    \centering
    \begin{tabular}{|*{1}{C{4cm}}| *{4}{C{3.0cm}}|}\hline
        GRB name & Angular separation & Loc. Uncertainty & Energy proxy & Time delay (s) \\ \hline
        GRB 130131B & 10.3$^\circ$ & 2.6$^\circ$ & 676 GeV & 54.0 s \\
        GRB 141220A & 2.0$^\circ$ & 2.2$^\circ$ & 47 GeV & 247.3 s \\
        GRB 160314B & 5.8$^\circ$ & 1.2$^\circ$ & 1023 GeV & 158.4 s \\
        GRB 160705B & 5.2$^\circ$ & 1.5$^\circ$ & 794 GeV & 91.5 s \\ 
        GRB 160912A & 6.1$^\circ$ & 2.3$^\circ$ & 525 GeV & 106.4 s \\ \hline
    \end{tabular}
    \label{tab:res_Prec_stack}
\end{table}

\section{Interpretation}\label{sec:results}
Since none of the analyses found evidence for neutrino emission from GRBs, limits can be placed on the GRB neutrino flux. These limits can be compared to model predictions and can constrain the fraction that GRBs contribute to the diffuse neutrino flux measured by IceCube~\citep{ICRC_diffuse}. It is worth noting that upper limits on the diffuse flux presented in previous IceCube publications \citep{IceCubeGRB2012,IceCubeGRB2015,IceCubeGRB2016,IceCubeGRB2017} correspond to the flux from GRBs observable by current gamma-ray satellites. In contrast, for the results presented here the limits are placed on the total contribution to the diffuse flux from all GRBs in the Universe. For the "Extended TW" and "Stacked precursor" analyses, the upper limits on the time integrated flux from all analysed GRBs, $\frac{dN}{dEdA}$, are converted to a diffuse flux
\begin{equation}\label{eq:diffuse_flux}
\Phi_\nu = \frac{dN}{dtd\Omega dEdA} = \epsilon_z \cdot \epsilon_d \cdot \frac{1}{\Delta t} \cdot \frac{1}{\Delta \Omega} \cdot \frac{dN}{dEdA}\ ,
\end{equation}
where the factor $\Delta \Omega$ normalises the flux by the solid angle subtended by the GRBs, i.e. $\Delta \Omega=4\pi$ for an all-sky sample, $\Delta t\sim7.16$~yr is the livetime of the IceCube data used in this work, $\epsilon_d$ corrects for the field-of-view (FOV) and dead-time of the GRB telescopes, and $\epsilon_z$ accounts for the contribution from GRBs that are too dim to be observable with current gamma-ray satellites. Determining the value of $\epsilon_d$ and $\epsilon_z$ requires specifying concrete characteristics of the GRB satellites. In the ``Extended TW'' analysis, which uses GRBs detected by a variety of satellites, the canonical estimate is made that with no dead-time and a all-sky FOV, a total of 667~GRBs would be observed every year \citep{IceCubeGRB2017}. In the ``Stacked precursor'' analysis, only well-localized GRBs are used, most of which were detected by \emph{Swift}. Limits are therefore set on the time-integrated flux of the subset of \emph{Swift} bursts, and then corrected using Eq.~\eqref{eq:diffuse_flux} based on the characteristics of the \emph{Swift} telescope. In the ``Precursor/Afterglow'' analysis, instead of finding a limit on the stacked flux and using it in Eq.~\eqref{eq:diffuse_flux}, a simulation of the neutrinos from individual GRBs following the cosmological distribution is used to provide simulated data sets for the analysis, allowing to set limits directly on the total flux by varying the injected signal strength from the population.

\subsection{Extended TW}\label{liz_results}
The stacked test statistic presented in Section \ref{sec:llh} is used to place an upper limit on contributions of GRBs to the quasi-diffuse flux measured by IceCube.  Flux is injected using an E$^{-2.28}$ \citep{ICRC_diffuse} spectrum until 90\% of trials yield a stacked test statistic above the unblinded value.  This injected flux is converted to a diffuse flux using Eq. \eqref{eq:diffuse_flux}.  This procedure is repeated for all ten time windows and the prompt.  For the prompt, the shortest time window that includes the entire reported $T_{100}$ is used (see Figure \ref{extended_tw_cartoon}).

\begin{figure}
    \centering
    \includegraphics[width=0.6\linewidth]{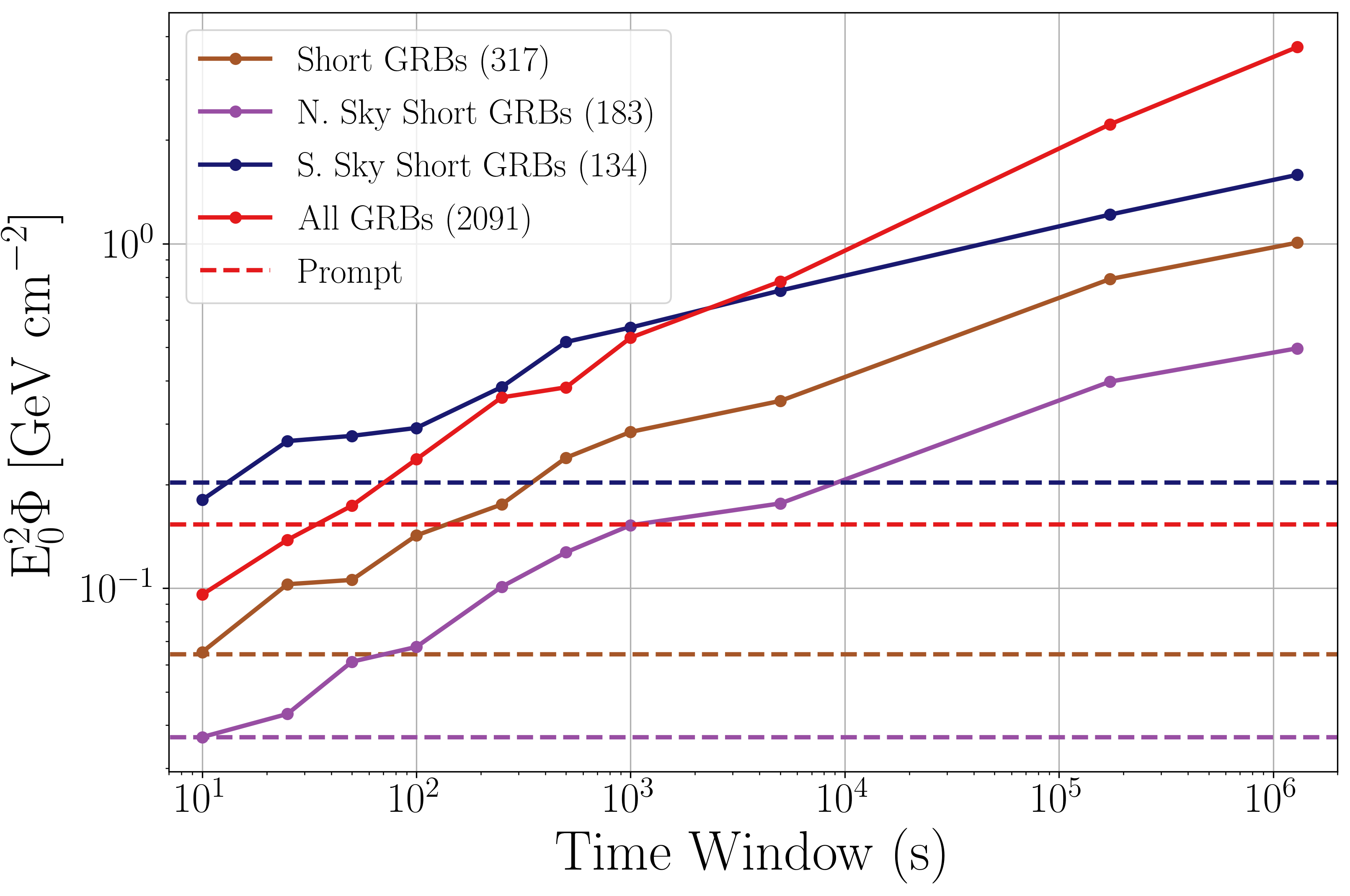}
    \caption{The time-integrated flux (at 1~TeV) for all short GRBs (dark blue) and the short GRBs split by Northern and Southern sky (red and light blue).  The limits for all 2,091 GRBs are shown in green for reference. Each dot indicates the 90\% confidence limit for the given time window, and the dashed lines show the limit for the prompt.}
    \label{fig:stacked_subpops}
\end{figure}

\begin{figure}
    \centering
    \includegraphics[width=0.45\linewidth]{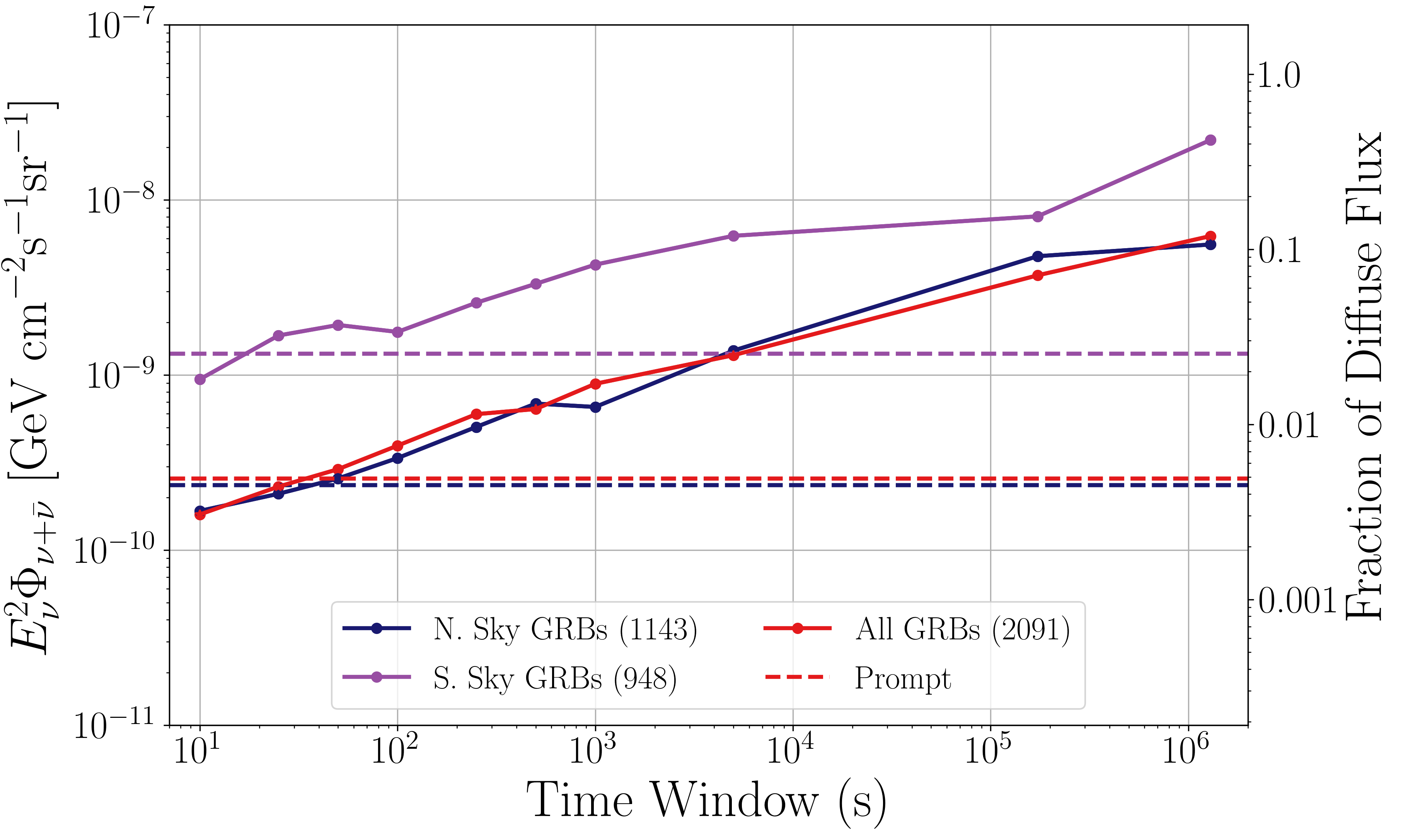}
    \includegraphics[width=0.45\linewidth]{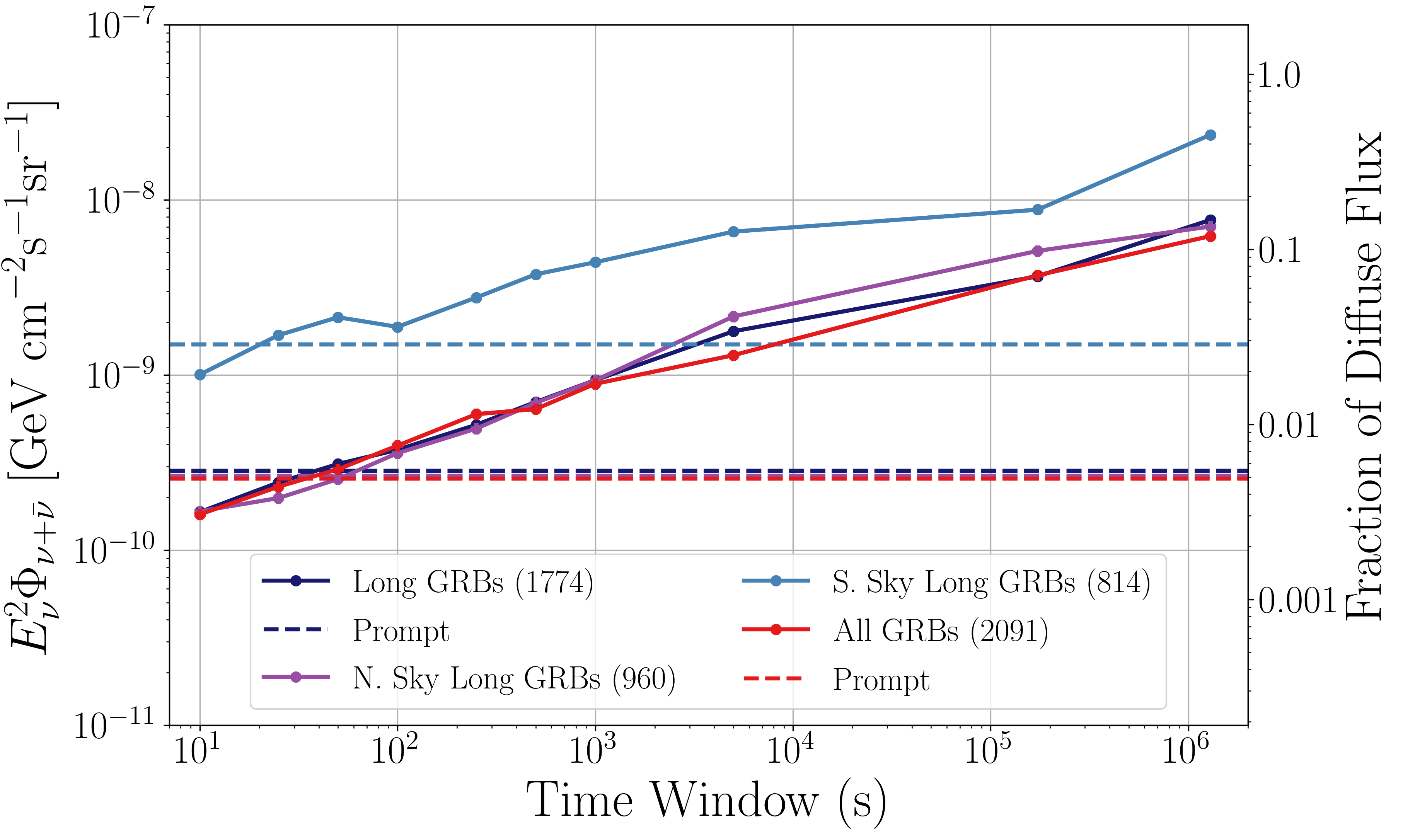}
    \caption{Left: The stacked limit on the quasi-diffuse flux (at 1~TeV) for all Northern GRBs (blue) and all Southern GRBs (red).  Right: The stacked limit on the quasi-diffuse flux for all long GRBs (dark blue) and the long GRBs split by northern and southern sky (red and light blue). In both figures, the right axis presents this limit as a fraction of the quasi-diffuse flux \citep{ICRC_diffuse}.  The limits for all 2,091 GRBs are shown in green for reference. Each dot indicates the 90\% confidence limit for the given time window, and the dashed lines show the limit for the prompt.}
    \label{fig:stacked_north_south_short_long}
\end{figure}

These 90\% confidence level limits are set for each time window for various subsets of GRBs.  The four sub-populations analyzed with a binomial test are presented in Figure \ref{fig:stacked_subpops}.  Limits are also placed on all GRBs observed in the Northern and Southern Sky, as well as all short and long GRBs (Figure \ref{fig:stacked_north_south_short_long}).  In each plot, the stacked limit from all 2,091 GRBs is shown for reference.

Previous IceCube studies \citep{IceCubeGRB2017} constrained the prompt contribution of GRBs observable by current gamma-ray satellites to $\sim$1\% of the diffuse flux observed by IceCube. The prompt limit presented here applies to all GRBs in the Universe and corresponds to $\lesssim$1\%.  The limits are similar despite analyzing nearly twice as many GRBs in this analysis.  The difference is the inclusion of $\epsilon_z$ term to account for GRBs that are too far away to be observable with current gamma-ray satellites. Given its fluence trigger threshold of $\sim10^{-8}$~erg~cm$^{-2}$~s$^{-1}$, the Swift-BAT detector can be assumed to view all canonical GRBs with an isotropic equivalent luminosity $L_{iso}\geq 10^{50}$~erg~cm$^{-2}$ up to a redshift of $z\sim1.3$. Assuming that all GRBs have identical properties in terms of neutrino emission and that they follow the redshift evolution described by~\cite{Swift_GRB}, the contribution from GRBs outside the observable redshift threshold can be calculated using the procedure outlined in~\citep{Ksi}. This results in $\epsilon_z$ value of $\sim2.1$. The inclusion of this $\epsilon_z$ term leads to a similarly constraining prompt limit compared to previous studies.

\subsection{Precursor/Afterglow}\label{Kunal}

\begin{figure}
    \centering
    \includegraphics[width=0.45\linewidth]{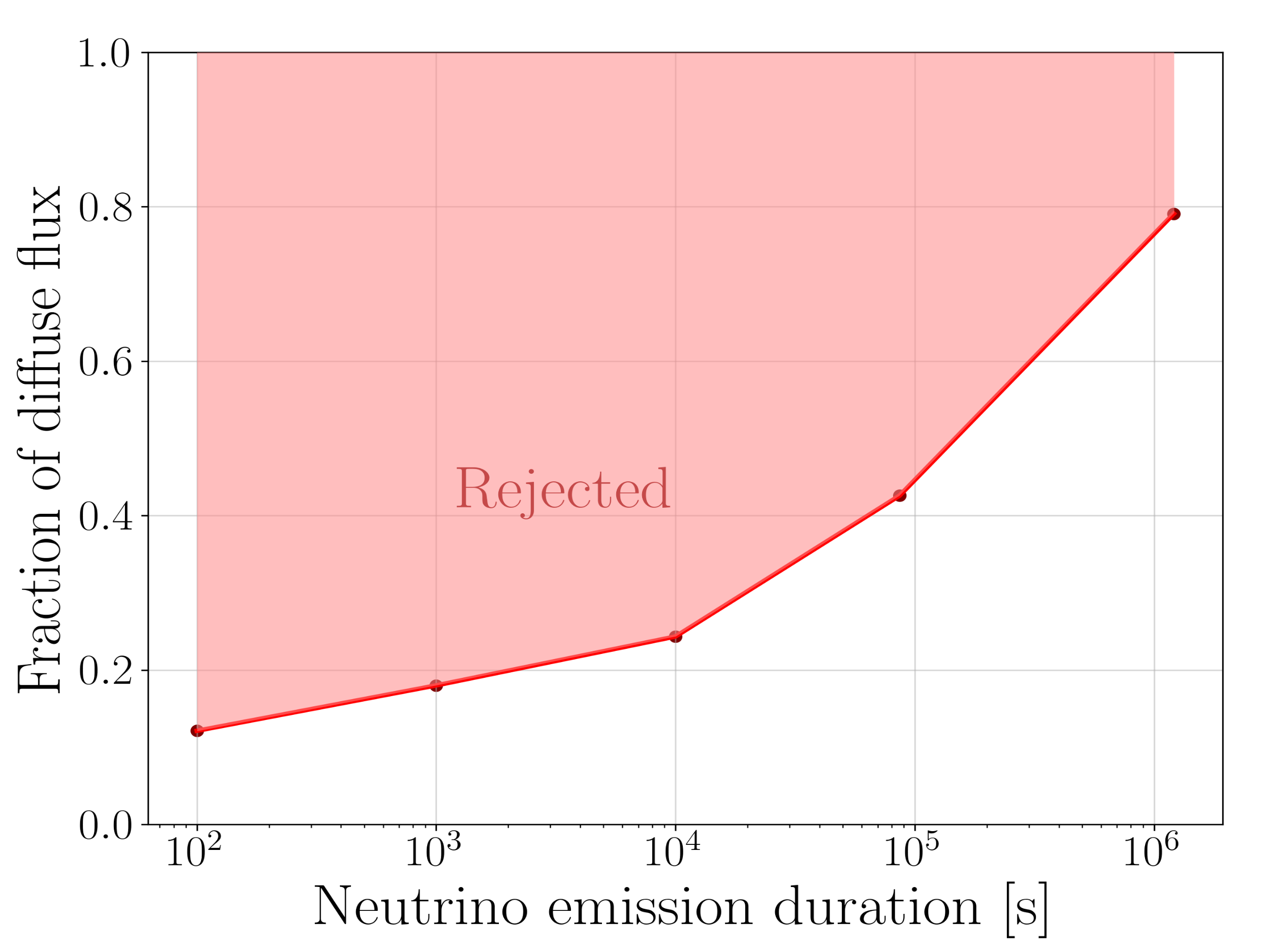}
    \caption{90\% CL upper-limit on the contribution to the quasi-diffuse neutrino flux~\citep{ICRC_diffuse} from the total population of all long GRBs~\cite{Swift_GRB}, for different durations of extended neutrino emission.}
    \label{fig:Limits_Kunal}
\end{figure}

These results were used to calculate limits for the total contribution of all long GRBs to the observed diffuse neutrino flux (see Figure~\ref{fig:Limits_Kunal}). Based on the study by~\cite{Swift_GRB} using Swift observations of GRBs to extrapolate to the whole Universe, a mean rate of 4,571 long GRBs per year is estimated to be beamed at Earth. We use the software package FIRESONG~\citep{Tung2021} to simulate neutrino emission from this cosmic GRB population, with different emission periods. Emission windows ranging from 100~s to 14~days were considered in the simulations. The redshift distribution of the cosmic GRB population from the study by~\cite{Swift_GRB} was used to simulate the GRB rate density in FIRESONG. We assume the case where no luminosity evolution is required and for simplicity assume that the GRBs are standard candle neutrino sources.

The Precursor/Afterglow analysis only considers 733 GRBs across 7.16 years, out of which 546 GRBs are long GRBs which were observed by~\emph{Swift}. Hence we use the GRB detection efficiency as a function of redshift from the study by~\cite{Swift_GRB} as well as the sky coverage and the survey time of ~\emph{Swift}/BAT to down-select from the ~4,571 sources created by the FIRESONG simulations to a sample of 546 GRBs to recreate the observation biases of the Swift sample. These 546 flux values down-selected from the simulation are used to inject signal with an $E^{-2.28}$~\citep{ICRC_diffuse} spectrum into a simulated neutrino dataset. To repeat the original analysis under the same conditions, an additional 187 sources are added with no signal (i.e. represent background only), bringing the sample to 733 sources again. The likelihood analysis and binomial test are then performed as before on this simulated sample. This sequence of steps for every emission window considered is repeated using different simulated neutrino datasets to produce 1000 trials.  When GRBs are assumed to produce the entire diffuse flux~\citep{ICRC_diffuse}, 100\% of these injected trials result in a binomial test result more significant than the unblinded result.  The fraction of the diffuse flux is then reduced to identify the flux where 90\% of trials produce a binomial test result more significant than the unblinded result.  This is summarized in Figure~\ref{fig:Limits_Kunal}, where the points show this upper limit (90\% confidence level) for a range of neutrino emission time windows. The respective fraction for the different emission durations considered represents the total allowed contribution of all long GRBs to the observed astrophysical diffuse neutrino flux. 

\subsection{Stacked Precursor}\label{StackedPrecursorRes}

\begin{figure}[t]
    \centering
    \includegraphics[width=0.60\linewidth]{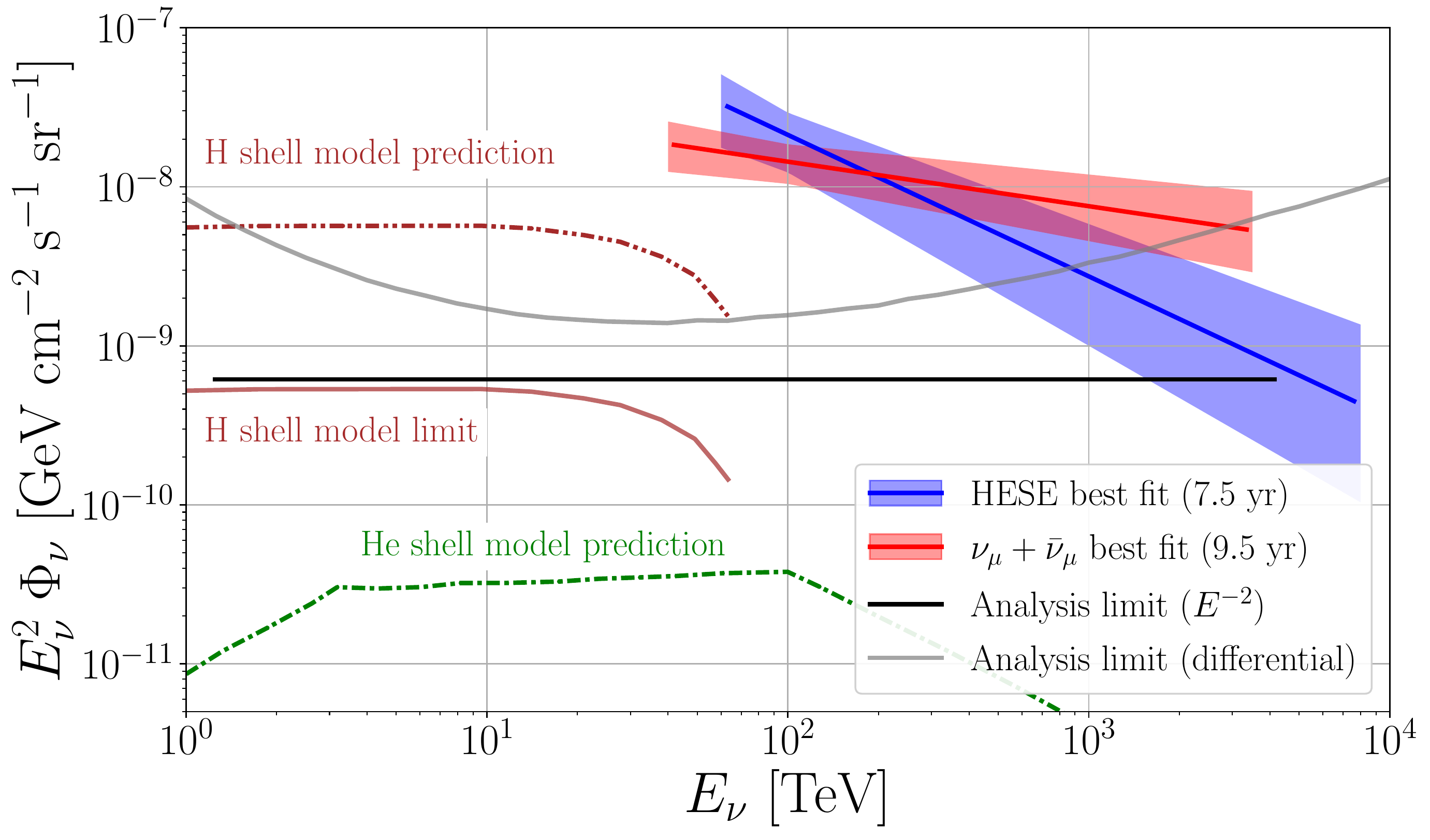}
    \caption{Model predictions by \citep{PrecursorModel} are shown for two progenitor scenarios. The red and green line corresponds to central engines enveloped by an outer hydrogen and helium shell, respectively. The ``Stacked precursor'' analysis can exclude the H-shell model by a factor 10, but does not constrain the He-shell model. Aside from these model comparisons, generic flux upper limits are also shown. Limits for an $E^{-2}$ spectrum are displayed by the black line, where the $x$-range corresponds to the energy band that contributes 90\% of all signal events. The solid grey line shows the differential flux, normalized per decade of energy, to which the analysis is sensitive. These generic upper limits extrapolate the total GRB flux in a different way than the model prediction/limits, as explained in \autoref{StackedPrecursorRes}. For reference, the astrophysical neutrino flux observed by IceCube is also shown \citep{IceCubeHESE75,ICRC_diffuse}.}
    \label{fig:Limits_Prec_stacking}
\end{figure}

Precursor neutrinos have been predicted in models in which a jet initially has to burrow its way through remnant layers of the progenitor star. A prediction for the diffuse precursor neutrino flux from such sources was made by \citep{PrecursorModel} and is shown in Figure \ref{fig:Limits_Prec_stacking}. The red and green lines correspond to progenitors that have a remnant outer hydrogen (H) or helium (He) shell, respectively. This analysis is able to exclude the H-shell model by a factor 10, but cannot constrain the He-shell model. To be consistent with the model prediction, the H-shell upper limit shown in Figure \ref{fig:Limits_Prec_stacking} assumes that the diffuse GRB neutrino flux results from $10^3$ GRBs per year \citep{PrecursorModel}. This is in contrast to the model-independent upper limits in Figure \ref{fig:Limits_Prec_stacking}, which rely on the conversion outlined in Eq. \eqref{eq:diffuse_flux}. This latter approach incorporates updated information about the redshift distribution of GRBs that was unavailable when the model was released. As a result, these generic limits are slightly more conservative.

\section{Conclusion}
The results from the four analyses presented in this paper cover 2209 unique GRBs. These GRBs are investigated for neutrino correlations from the precursor, prompt, and afterglow emission regions in a comprehensive manner and all the four analyses report observations consistent with background expectations. We obtain a range of upper limits to the astrophysical diffuse neutrino flux. We show that prompt emission from all GRBs in the Universe is limited to 1\% of the diffuse astrophysical neutrino flux. Neutrino emission limits range from 1\%-2\% in timescales up to $10^3$ s using the historic assumption of 667 GRBs observable by satellites per year. Neutrino emission is constrained to less than 24\%  for timescales up to $10^4$ s by simulating GRB populations using the FIRESONG~\citep{Tung2021} module. These constraints are shown for additional neutrino emission timescales in Figures~\ref{fig:stacked_subpops}, \ref{fig:stacked_north_south_short_long} and \ref{fig:Limits_Kunal}. By looking for neutrinos on time scales motivated by the observation of gamma-ray precursors, we were able to constrain physical models such as those presented by~\cite{PrecursorModel} (see Figure~\ref{fig:Limits_Prec_stacking}).
 
Table~\ref{table:HESS_GRB} highlights the result for the GRB 180720B which was observed by H.E.S.S. to have a high-energy afterglow counterpart 10 hours after the start of prompt emission~\citep{328}. The exceptionally bright GRB~130427A~\citep{Bright_GRB} was also analyzed, but none of the four analyses reported neutrino excesses in correlation with it. The results for the most significant GRBs from the individual analyses are presented in Appendix \ref{AppendixA} and \ref{AppendixB}. A comprehensive summary of all the results from each analysis can be accessed from the supplementary materials. 

\begin{table}[H]
    \centering
    \caption{Summary of GRB information and best-fit results for the GRB 180720B. The positional coordinates are defined using the equatorial coordinates: right ascension ($\alpha$) and declination ($\delta$). These two quantities are expressed in degrees. $T_{100}$ represents the total period of observation of prompt $\gamma$ fluence from the GRB and is expressed in seconds. $\hat{n_s}$ represents the best-fit number of signal events, $\hat{\gamma}$ represents best-fit spectral index and $\hat{T}_w$ is the best-fit time window (expressed in days). $E^{2}F$ is the 90\% confidence level upper limit on the time integrated flux normalization at 1 TeV in GeV/$\text{cm}^{2}$. For Precursor results, the limit is evaluated for the best fit values of $\hat{\gamma}$ and $\hat{T}_w$, and for the Afterglow results $\gamma$=2.0 and ${T}_w$=12.1~hrs is used to match with the H.E.S.S observation period~\citep{328}. For GRB 130427A, $\gamma$=2.0 was considered for all the $E^{2}F$ calculations. For Extended TW, the limit is evaluated for the prompt phase, for Precursor/Afterglow ${T}_w$=14.0~days is used. }
    \label{table:HESS_GRB}
    \begin{tabular}{|c c c c|c c c c|c c c c c|c c|}
        \hline
        \multicolumn{4}{|c|}{GRB information} & \multicolumn{4}{c|}{Extended TW}  & \multicolumn{5}{c|}{Precursor} &
        \multicolumn{2}{c|}{Afterglow}\\ 
        \hline
        GRB & $\alpha$ & $\delta$ & $T_{100}$ & TW & $\hat{n}_s$ & $E^{2}F$ & \textit{p}-value & $\hat{T}_w$ & $\hat{n}_s$ & $\hat{\gamma}$ & \textit{p}-value & $E^{2}F$ & \textit{p}-value & $E^{2}F$\\
        \hline
        180720B & 0.53 & $-2.92$ & 53.9 & $\pm$1 & 1.3 & 0.05 & 7.2e-02 & $-8.6$ & 3.6 & 2.32 & 1.5e-02 & 0.26 & 1.0 & 0.03 \\
        \hline
        130427A & 173.14 & 27.70 & 213.83 & - & - & 0.05 & 1.0 & - & - & - & 1.0 & 0.06 & 1.0 &  0.06 \\  
        \hline
    \end{tabular}
\end{table}

Despite non-detections in these analyses, GRBs cannot be fully eliminated as potential sources of high-energy neutrinos.  A larger class of neutrino detector, such as the proposed IceCube-Gen2 detector, will offer increased sensitivity to observe potential neutrino-emitting sub-populations of GRBs.

\clearpage

\section*{Acknowledgements}
The IceCube Collaboration acknowledges the significant contribution to this manuscript from Paul Coppin, Kunal Deoskar and Elizabeth Friedman. The authors thank Adam Goldstein for providing GBM maps prior to 2018 and gratefully acknowledge NASA funding for this work through contract NNM13AA43C. Joshua Wood gratefully acknowledges NASA funding through the Fermi-GBM project. The authors gratefully acknowledge the support from the following agencies and institutions: USA {\textendash} U.S. National Science Foundation-Office of Polar Programs,
U.S. National Science Foundation-Physics Division,
U.S. National Science Foundation-EPSCoR,
Wisconsin Alumni Research Foundation,
Center for High Throughput Computing (CHTC) at the University of Wisconsin{\textendash}Madison,
Open Science Grid (OSG),
Extreme Science and Engineering Discovery Environment (XSEDE),
Frontera computing project at the Texas Advanced Computing Center,
U.S. Department of Energy-National Energy Research Scientific Computing Center,
Particle astrophysics research computing center at the University of Maryland,
Institute for Cyber-Enabled Research at Michigan State University,
and Astroparticle physics computational facility at Marquette University;
Belgium {\textendash} Funds for Scientific Research (FRS-FNRS and FWO),
FWO Odysseus and Big Science programmes,
and Belgian Federal Science Policy Office (Belspo);
Germany {\textendash} Bundesministerium f{\"u}r Bildung und Forschung (BMBF),
Deutsche Forschungsgemeinschaft (DFG),
Helmholtz Alliance for Astroparticle Physics (HAP),
Initiative and Networking Fund of the Helmholtz Association,
Deutsches Elektronen Synchrotron (DESY),
and High Performance Computing cluster of the RWTH Aachen;
Sweden {\textendash} Swedish Research Council,
Swedish Polar Research Secretariat,
Swedish National Infrastructure for Computing (SNIC),
and Knut and Alice Wallenberg Foundation;
Australia {\textendash} Australian Research Council;
Canada {\textendash} Natural Sciences and Engineering Research Council of Canada,
Calcul Qu{\'e}bec, Compute Ontario, Canada Foundation for Innovation, WestGrid, and Compute Canada;
Denmark {\textendash} Villum Fonden and Carlsberg Foundation;
New Zealand {\textendash} Marsden Fund;
Japan {\textendash} Japan Society for Promotion of Science (JSPS)
and Institute for Global Prominent Research (IGPR) of Chiba University;
Korea {\textendash} National Research Foundation of Korea (NRF);
Switzerland {\textendash} Swiss National Science Foundation (SNSF);
United Kingdom {\textendash} Department of Physics, University of Oxford.

\bibliography{references}

\appendix

\section{Extended TW: Most Significant GRBs}\label{AppendixA}


This appendix provides details for the most significant GRBs in the Extended TW analysis.  The threshold for inclusion in these tables is a \textit{p}-value below 1\%.  For the case of short GRBs detected in the southern hemisphere, a threshold of 5\% was chosen to include five GRBs in Table \ref{table:south_short}.  This was done to match the results of the binomial test, which indicated the most significant subset of the southern sky short GRBs was $k$=5.  The other sub-populations had a much larger value of $k$ and would have increased the \textit{p}-value threshold beyond a reasonable cutoff.  Although the results of the binomial test were consistent with background, these additional GRBs are included in Table \ref{table:south_short} for completeness.

\subsection{Variables}\label{appendix_A_variables}
The GRB name is based on GCN notices, with a preference for the name provided by Fermi-GBM.  If the GRB is not observed by Fermi-GBM, then the name is based on the date of observation with the format YYMMDD.  The final letter indicates the order of detection if more than one GRB is observed in a single day.  The right ascension ($\alpha$) and declination ($\delta$) in J2000.0 equatorial coordinates, as well as the localization uncertainty ($\sigma$) are all listed in degrees.  $T_{0}$ indicates the MJD trigger time of the GRB.  The $T_{100}$ is provided in seconds.

The most significant time window selected by this analysis is indicated by ``TW" and is given in either seconds or days with units provided in the column.  The test statistic (TS) and best fit number of signal events ($\hat{n}_s$) are provided as well as the \textit{p}-value in the final column.  The \textit{p}-value has an effective trial correction for searching 10 time windows, but it is not corrected for the size of the given sub-population.

\subsection{Results by Sub-Population}\label{appendix_A_tables}
The GRBs are split into sub-populations by duration and hemisphere.  Short GRBs are defined as $T_{90}<2~s$ and long GRBs are $T_{90}\ge2~s$.  In this analysis, the northern sky is defined as a declination greater than $-5^\circ$, while the southern sky includes all declinations less than $-5^\circ$.  The most significant GRBs for each sub-population are shown in the following tables.  The information is split into two sections: GRB information based on GCN notices and Fit results based on the analysis.

\begin{table}[H]\label{}
    \centering
    \caption{The most significant long GRBs from the northern hemisphere.  This sub-population has a total of 960 GRBs.  The variables are described in detail in \ref{appendix_A_variables}.}
    \begin{tabular}{| *{1}{p{2.35cm}} *{2}{p{1.15cm}} *{1}{p{1.5cm}} *{1}{p{1.65cm}} *{1}{p{0.5cm}} *{1}{p{1.0cm}}| *{3}{p{1.0cm}} *{1}{p{1.5cm}} |}
        \hline
        \multicolumn{7}{|c|}{GRB information} & \multicolumn{4}{|c|}{Fit results} \\ 
        \hline
       \centering GRB Name & \centering $\alpha$ & \centering $\delta$ & \centering $\sigma$ & \centering $T_{0}$ & \ $z$& $T_{100}$ & TW & \ TS & \ $\hat{n}_s$ & \ \textit{p}-value \\
        \hline
        GRB 140607A & 86.37 & 18.90 & 1.48e-02 & 56815.718 & --- & 109.90 & 2 d & 12.80 & 1.00 & 6.00e-04 \\ 
        \hline
        GRB 141121A & 122.67 & 22.22 & 6.47e-05 & 56982.150 & 1.47 & 1419.90 & 15 d & 11.85 & 1.23 & 1.70e-03 \\
        \hline
        GRB 161125931 & 59.36 & 28.13 & 4.65e+00 & 57717.930 & --- & 69.63 & 25 s & 18.75 & 1.00 & 2.10e-03 \\ 
        \hline
        GRB 120911A & 357.98 & 63.10 & 2.07e-04 & 56181.298 & --- & 22.02 & 250 s & 10.62 & 1.00 & 2.50e-03 \\ 
        \hline
        GRB 120504468 & 329.94 & 46.83 & 8.76e+00 & 56051.468 & --- & 41.98 & 1000 s & 19.41 & 1.00 & 3.60e-03 \\ 
        \hline
        GRB 170131A & 341.45 & 64.01 & 2.33e-02 & 57784.969 & --- & 23.04 & 5000 s & 10.57 & 1.00 & 4.10e-03 \\ 
        \hline
        GRB 120711B & 331.69 & 60.02 & 2.33e-04 & 56119.133 & --- & 60.00 & 2 d & 11.83 & 3.21 & 4.20e-03 \\ 
        \hline
        GRB 180721A & 347.71 & 4.86 & 2.85e-04 & 58320.463 & --- & 47.60 & 2 d & 11.69 & 2.22 & 4.30e-03 \\ 
        \hline
        GRB 140114A & 188.52 & 27.95 & 1.81e-04 & 56671.498 & 3.0 & 139.70 & 2 d & 10.99 & 1.19 & 7.50e-03 \\ 
        \hline
        GRB 160201883 & 312.67 & 69.32 & 3.14e+00 & 57419.883 & --- & 40.51 & 1000 s & 13.42 & 1.96 & 7.50e-03 \\ 
        \hline
        GRB 120217808 & 122.44 & 36.77 & 5.03e+00 & 55974.808 & --- & 5.89 & 25 s & 15.75 & 1.00 & 7.70e-03 \\ 
        \hline
        GRB 180623849 & 199.40 & -4.26 & 2.52e+00 & 58292.849 & --- & 17.73 & 50 s & 11.48 & 1.00 & 9.60e-03 \\ 
        \hline
    \end{tabular}
\end{table}
\begin{table}[H]\label{}
    \centering
    \caption{The most significant short GRBs from the northern hemisphere.  This sub-population has a total of 183 GRBs. The variables are described in detail in \ref{appendix_A_variables}.}
    \begin{tabular}{| *{1}{p{2.35cm}} *{2}{p{1.15cm}} *{1}{p{1.5cm}} *{1}{p{1.65cm}} *{1}{p{0.5cm}} *{1}{p{1.0cm}}| *{3}{p{1.0cm}} *{1}{p{1.5cm}} |}
        \hline
        \multicolumn{7}{|c|}{GRB information} & \multicolumn{4}{|c|}{Fit results} \\ 
        \hline
       \centering GRB Name & \centering $\alpha$ & \centering $\delta$ & \centering $\sigma$ & \centering $T_{0}$ & \ $z$& $T_{100}$ & TW & \ TS & \ $\hat{n}_s$ & \ \textit{p}-value \\
        \hline
        GRB 140807500 & 200.16 & 26.49 & 4.39e+00 & 56876.500 & --- & 0.51 & 100 s & 15.72 & 1.00 & 4.80e-03 \\ 
        \hline
    \end{tabular}
\end{table}
\begin{table}[H]\label{}
    \centering
    \caption{The most significant long GRBs from the southern hemisphere.  This sub-population has a total of 814 GRBs.  The variables are described in detail in \ref{appendix_A_variables}.}
    \begin{tabular}{| *{1}{p{2.35cm}} *{2}{p{1.15cm}} *{1}{p{1.5cm}} *{1}{p{1.65cm}} *{1}{p{0.5cm}} *{1}{p{1.0cm}}| *{3}{p{1.0cm}} *{1}{p{1.5cm}} |}
        \hline
        \multicolumn{7}{|c|}{GRB information} & \multicolumn{4}{|c|}{Fit results} \\ 
        \hline
       \centering GRB Name & \centering $\alpha$ & \centering $\delta$ & \centering $\sigma$ & \centering $T_{0}$ & \ $z$& $T_{100}$ & TW & \ TS & \ $\hat{n}_s$ & \ \textit{p}-value \\
        \hline
        GRB 150202A & 39.23 & -33.15 & 2.20e-04 & 57055.965 & --- & 25.70 & 2 d & 18.61 & 2.93 & 5.00e-04 \\ 
        \hline
        GRB 150118B & 240.24 & -35.75 & 5.00e-01 & 57040.409 & --- & 48.65 & 2 d & 12.86 & 1.67 & 3.00e-03 \\ 
        \hline
        GRB 170923566 & 228.30 & -10.78 & 9.05e+00 & 58019.566 & --- & 27.65 & 1000 s & 19.72 & 1.00 & 4.10e-03 \\ 
        \hline
        GRB 150626A & 111.34 & -37.78 & 2.33e-04 & 57199.092 & --- & 144.00 & 15 d & 11.22 & 3.13 & 7.40e-03 \\ 
        \hline
        GRB 180906597 & 104.81 & -67.02 & 3.62e+00 & 58367.597 & --- & 52.03 & 25 s & 9.72 & 1.00 & 8.60e-03 \\ 
        \hline
    \end{tabular}
\end{table}
\begin{table}[H]\label{table:south_short}
    \centering
    \caption{The most significant short GRBs from the southern hemisphere.  This sub-population has a total of 134 GRBs.  The variables are described in detail in \ref{appendix_A_variables}.}
    \begin{tabular}{| *{1}{p{2.35cm}} *{2}{p{1.15cm}} *{1}{p{1.5cm}} *{1}{p{1.65cm}} *{1}{p{0.5cm}} *{1}{p{1.0cm}}| *{3}{p{1.0cm}} *{1}{p{1.5cm}} |}
        \hline
        \multicolumn{7}{|c|}{GRB information} & \multicolumn{4}{|c|}{Fit results} \\ 
        \hline
       \centering GRB Name & \centering $\alpha$ & \centering $\delta$ & \centering $\sigma$ & \centering $T_{0}$ & \ $z$& $T_{100}$ & TW & \ TS & \ $\hat{n}_s$ & \ \textit{p}-value \\
        \hline
        GRB 140511095 & 329.76 & -30.06 & 8.50e+00 & 56788.095 & --- & 1.41 & 2 d & 20.27 & 2.93 & 9.20e-03 \\ 
        \hline
        GRB 130919173 & 297.35 & -11.73 & 5.23e+00 & 56554.173 & --- & 0.96 & 5000 s & 16.20 & 1.00 & 1.75e-02 \\ 
        \hline
        GRB 160411A & 349.36 & -40.24 & 2.85e-04 & 57489.062 & --- & 1.26 & 2 d & 8.08 & 0.99 & 2.36e-02 \\ 
        \hline
        GRB 120212353 & 303.40 & -48.10 & 1.05e+01 & 55969.353 & --- & 0.86 & 500 s & 17.82 & 1.00 & 2.78e-02 \\ 
        \hline
        GRB 141102112 & 223.23 & -17.42 & 9.21e+00 & 56963.112 & --- & 0.02 & 15 d & 18.56 & 2.97 & 2.81e-02 \\ 
        \hline
    \end{tabular}
\end{table}

\section{Precursor/Afterglow: Top 20 GRBs}\label{AppendixB}


\begin{table}\label{table:Precursor}
    \centering
    \caption{Summary of best fit results and information for the GRBs in the top 20 results for the Precursor search. Each GRB is named based on the date when it was observed with the standard formatting of YYMMDD and a letter denoting the order in which the bursts were detected on the same day (A, B etc.). The positional coordinates for the bursts are defined using the equatorial coordinates: right ascension ($\alpha$) and declination ($\delta$). These two quantities are expressed in degrees. The burst timing ($T_0$) is expressed using the dating convention of Modified Julian Date (MJD). $z$ represents the redshift. $T_{100}$ represents the total period of the GRB observation and is expressed in seconds. $\hat{n_s}$ represents the best-fit number of signal events ${n_s}$, $\hat{\gamma}$ represents best-fit spectral index and $\hat{T}_w$ is the best-fit time window (expressed in seconds). `TS' represents the test statistic and `Significance' represents the significance of the pre-trial local \textit{p}-value for one-sided Gaussian distributions.}{\label{result_table_PC}}
    \begin{tabular}{|c c c c c c | c c c c c c|}
        \hline
        \multicolumn{6}{|c|}{GRB information} & \multicolumn{6}{|c|}{Fit results}\\
        \hline
        GRB Name & $\alpha$ & $\delta$ & $T_0$ & $z$ & $T_{100}$ & $\hat{n}_s$ & $\hat{\gamma}$ & $\hat{T}_w$ & TS & \textit{p}-value & Significance \\
        \hline
        GRB 150202A & 39.23 & -33.15 & 57055.965301 & -- & 25.70 & 1.00 & 4.00 & 3.367e+03 & 16.37 & 6.12e-04 & 3.23$\sigma$ \\
        GRB 180721A & 347.71 & 4.86 & 58320.463056 & -- & 47.60 & 1.00 & 1.84 & 1.542e+04 & 12.46 & 2.73e-03 & 2.78$\sigma$ \\
        GRB 140301A & 69.56 & -34.26 & 56717.642234 & 1.42 & 31.00 & 1.96 & 2.15 & 7.615e+05 & 11.51 & 4.38e-03 & 2.62$\sigma$ \\
        GRB 141220A & 195.07 & 32.15 & 57011.251986 & 1.32 & 7.62 & 1.00 & 4.00 & 2.473e+02 & 11.19 & 5.39e-03 & 2.55$\sigma$ \\
        GRB 111126A & 276.06 & 51.46 & 55891.790069 & -- & 0.80 & 1.84 & 4.00 & 3.556e+03 & 10.65 & 6.22e-03 & 2.50$\sigma$ \\
        GRB 151205A & 229.29 & 35.74 & 57361.656944 & -- & 62.80 & 1.00 & 3.80 & 6.390e+03 & 10.10 & 8.15e-03 & 2.40$\sigma$ \\
        GRB 170531B & 286.88 & -16.42 & 57904.918160 & 2.37 & 164.13 & 2.70 & 2.59 & 5.077e+05 & 9.35 & 9.17e-03 & 2.36$\sigma$ \\
        GRB 171007A & 135.60 & 42.82 & 58033.498356 & -- & 105.00 & 2.63 & 2.47 & 2.963e+04 & 9.72 & 9.94e-03 & 2.33$\sigma$ \\
        GRB 160310A & 98.82 & -7.22 & 57457.015943 & -- & 26.60 & 0.99 & 1.83 & 4.455e+04 & 8.59 & 1.19e-02 & 2.26$\sigma$ \\
        GRB 180720B & 0.53 & -2.92 & 58319.598368 & 0.65 & 53.90 & 3.59 & 2.32 & 7.435e+05 & 8.89 & 1.50e-02 & 2.17$\sigma$ \\
        GRB 160422A & 42.09 & -57.88 & 57500.499303 & -- & 14.12 & 0.99 & 2.12 & 4.246e+04 & 7.16 & 1.96e-02 & 2.06$\sigma$ \\
        GRB 140619A & 27.11 & -39.26 & 56827.485127 & -- & 233.90 & 0.98 & 2.54 & 8.499e+04 & 6.21 & 2.47e-02 & 1.96$\sigma$ \\
        GRB 160629A & 4.82 & 76.98 & 57568.930208 & 3.33 & 76.38 & 3.17 & 3.40 & 3.327e+05 & 7.06 & 2.54e-02 & 1.95$\sigma$ \\
        GRB 131014A & 100.30 & -19.10 & 56579.214583 & -- & 4.36 & 0.98 & 3.31 & 6.707e+03 & 6.38 & 2.60e-02 & 1.94$\sigma$ \\
        GRB 151027A & 272.49 & 61.35 & 57322.165556 & 0.81 & 129.69 & 0.99 & 2.90 & 4.185e+03 & 7.59 & 2.70e-02 & 1.93$\sigma$ \\
        GRB 131218A & 113.80 & -64.72 & 56644.878843 & -- & -- & 0.98 & 2.97 & 3.542e+05 & 5.84 & 2.79e-02 & 1.91$\sigma$ \\
        GRB 120722A & 230.50 & 13.25 & 56130.537106 & 0.96 & 42.40 & 2.42 & 2.51 & 3.223e+05 & 7.39 & 2.87e-02 & 1.90$\sigma$ \\
        GRB 120711B & 331.69 & 60.02 & 56119.132669 & -- & 60.00 & 2.52 & 2.59 & 7.865e+04 & 7.31 & 2.88e-02 & 1.90$\sigma$ \\
        GRB 150627A & 117.47 & -51.49 & 57200.182905 & -- & 70.57 & 0.98 & 1.73 & 8.081e+05 & 5.58 & 2.92e-02 & 1.89$\sigma$ \\
        GRB 131030A & 345.07 & -5.37 & 56595.872428 & 1.29 & 41.29 & 0.99 & 3.81 & 3.250e+03 & 7.07 & 2.98e-02 & 1.88$\sigma$ \\
        \hline
    \end{tabular}
\end{table}


\begin{table}[h]\label{table:Afterglow}
    \centering
    \caption{Summary of best fit results and information for the GRBs in the top 20 results for the Prompt+Afterglow search. This is similar to Table~\ref{result_table_PC} but shows results from the prompt+afterglow search instead. }{\label{result_table_AG}}
    \begin{tabular}{|c c c c c c | c c c c c c|}
        \hline
        \multicolumn{6}{|c|}{GRB information} & \multicolumn{6}{|c|}{Fit results}\\
        \hline 
        GRB Name & $\alpha$ & $\delta$ & $T_0$ & $z$ & $T_{100}$ & $\hat{n}_s$ & $\hat{\gamma}$ & $\hat{T}_w$ & TS & \textit{p}-value & Significance \\
        \hline
        GRB 170318A & 305.67 & 28.41 & 57830.508287  & -- & 133.70 & 2.91 & 3.52 & 4.267e+04 & 16.13 & 6.11e-04 & 3.23$\sigma$ \\
        GRB 140607A & 86.37 & 18.90 & 56815.717720  & -- & 109.90 & 1.00 & 1.53 & 2.602e+04 & 15.02 & 9.35e-04 & 3.11$\sigma$ \\
        GRB 141121A & 122.67 & 22.22 & 56982.160220 & 1.47 & 1419.90 & 1.17 & 1.38 & 1.040e+06 & 13.28 & 1.81e-03 & 2.91$\sigma$ \\
        GRB 140114A & 188.52 & 27.95 & 56671.498380 & 3.00 & 139.70 & 1.00 & 1.14 & 8.478e+04 & 12.20 & 3.62e-03 & 2.69$\sigma$ \\
        GRB 120911A & 357.98 & 63.10 & 56181.297564 & -- & 22.02 & 1.00 & 2.49 & 1.219e+02 & 11.77 & 3.86e-03 & 2.66$\sigma$ \\
        GRB 140930B & 6.35 & 24.29 & 56930.820625 & -- & 0.84 & 1.00 & 4.00 & 6.691e+03 & 10.34 & 8.08e-03 & 2.41$\sigma$ \\
        GRB 150317A & 138.98 & 55.47 & 57098.182431 & -- & 23.29 & 2.76 & 4.00 & 6.264e+04 & 9.79 & 9.12e-03 & 2.36$\sigma$ \\
        GRB 160827A & 179.27 & -29.18 & 57627.657465 & -- & 13.30 & 1.00 & 4.00 & 1.426e+05 & 8.91 & 9.12e-03 & 2.36$\sigma$ \\
        GRB 180418A & 170.12 & 24.93 & 58226.280625 & -- & 2.78 & 1.81 & 1.78 & 9.165e+04 & 9.85 & 9.65e-03 & 2.34$\sigma$ \\
        GRB 130313A & 236.41 & -0.37 & 56364.672350 & -- & 0.26 & 2.81 & 1.96 & 4.446e+05 & 9.51 & 1.29e-02 & 2.23$\sigma$ \\
        GRB 131202A & 344.05 & -21.66 & 56628.633409 & 7.50 & 32.90 & 0.99 & 4.00 & 1.315e+05 & 8.01 & 1.29e-02 & 2.23$\sigma$ \\
        GRB 170728B & 237.98 & 70.12 & 57962.960630 & -- & 48.29 & 1.90 & 2.41 & 1.080e+04 & 8.63 & 1.42e-02 & 2.19$\sigma$ \\
        GRB 170604A & 342.66 & -15.41 & 57908.797801 & 1.33 & 26.70 & 0.99 & 4.00 & 1.103e+05 & 7.80 & 1.52e-02 & 2.16$\sigma$ \\
        GRB 140730A & 56.40 & -66.55 & 56868.822118& -- & 41.30 & 0.99 & 3.94 & 2.110e+04 & 7.68 & 1.60e-02 & 2.14$\sigma$ \\
        GRB 160411A & 349.36 & -40.24 & 57489.061701 & -- & 1.26 & 0.99 & 2.79 & 4.832e+04 & 7.47 & 1.63e-02 & 2.14$\sigma$ \\
        GRB 150725A & 220.42 & -2.42 & 57228.364056 & -- & -- & 1.00 & 4.00 & 7.339e+02 & 8.86 & 1.72e-02 & 2.11$\sigma$ \\
        GRB 180823A & 210.36 & 14.89 & 58353.794815 & -- & 80.30 & 1.56 & 4.00 & 2.689e+04 & 8.42 & 1.94e-02 & 2.07$\sigma$ \\
        GRB 150912A & 248.43 & -20.98 & 57277.442708 & -- & 34.82 & 0.99 & 3.48 & 1.423e+05 & 6.65 & 2.04e-02 & 2.05$\sigma$ \\
        GRB 160221A & 232.08 & -28.45 & 57439.992847 & -- & 12.95 & 0.99 & 1.84 & 7.981e+04 & 6.47 & 2.20e-02 & 2.01$\sigma$ \\
        GRB 160424A & 319.49 & -60.41 & 57502.492429 & -- & 7.46 & 1.82 & 1.95 & 6.239e+05 & 6.49 & 2.35e-02 & 1.99$\sigma$ \\
        \hline
    \end{tabular}
\end{table}

\end{document}